\documentclass[traditabstract]{aa} 

\usepackage{graphicx}
\usepackage{txfonts}
\usepackage{url}
\usepackage{natbib}
\bibpunct{(}{)}{;}{a}{}{,} 
\usepackage[utf8]{inputenc}
\usepackage{subfigure}
\usepackage{verbatim}
\begin{document}

   \title{First evidence of interaction between longitudinal and transverse waves in solar magnetic elements}
   \author{M. Stangalini$^{1}$, S. K. Solanki$^{1,2}$, R. Cameron$^{1}$, V. Martinez Pillet$^{3,4}$}
   \institute{$^{1}$Max Planck Institute for Solar System Research, Max-Planck-Str. 2 37191 Katlenburg-Lindau, Germany\\
   $^{2}$School of Space Research, Kyung Hee University, Yongin, Gyeonggi 446-701, Republic of Korea\\
    $^{3}$ IAC Instituto de Astrof\'isica de Canarias, V\'ia L\'actea s/n, E-38205 La Laguna, Tenerife, Spain\\
    $^{4}$ Departamento de Astrof\'isica, Universidad de La Laguna, E-38205 La Laguna, Tenerife, Spain\\
              \email{stangalini@mps.mpg.de}}


  \abstract 
{Small-scale magnetic fields are thought to play an important role in the heating of the outer solar atmosphere. By taking advantage of the unprecedented high-spatial and temporal cadence of IMaX, the filter vector polarimeter on board the Sunrise balloon-borne observatory, we study the transversal and longitudinal velocity oscillations in small magnetic elements. The results of this analysis are then compared to MHD simulations, showing excellent agreement. We found buffeting-induced transverse oscillations with velocity amplitudes of the order of $1-2$ km/s, to be common along with longitudinal oscillations with amplitudes $\sim 0.4$ km/s. Moreover, we also found an interaction between transverse oscillations and longitudinal velocity oscillations, showing a $\pm 90^{\circ}$ phase lag at the frequency at which they exhibit the maximum coherence in the power spectrum. Our results are consistent with the theoretical picture in which MHD longitudinal waves are excited inside small magnetic elements as a response of the flux tube to the forcing action of the granular flows.
}  

   \authorrunning{M. Stangalini}
	\titlerunning{MHD waves in small magnetic features}
\maketitle

\section{Introduction}
Magnetic fields play a major role in the dynamics and the energetics of the solar atmosphere. They are present over a wide range of spatial scales \citep{2006RPPh...69..563S}, from very small elements \citep{2010ApJ...723L.164L}, at or below the best current spatial resolution achieved by modern solar telescopes ($\simeq 100$ km), up to large sunspots with diameters of order of $70000$ km. Interestingly, magnetic fields, organized in structures similar to flux tubes or sheets, can act as a wave-guide for waves and perturbations, through different layers of the solar atmosphere \citep{1996SSRv...75..453N, DePontieu2004}. It has been estimated that concentrated small-scale magnetic fields cover roughly $1\%$ of the quiet solar surface \citep{2012A&A...539A...6B}.\\
Much work has been done, both theoretically and through numerical simulations, to investigate and to model the wave propagation in slender magnetic flux tubes \citep[e.g.][to name a few]{1978SoPh...56....5R, Edwin1983, Roberts1983, Musielak1989, 1998ApJ...495..468S, Hasan2003, Musielak2003a, Khomenko2008, Fedun2011}. In contrast, comparatively few observational studies exist about waves in the abundant small-scale magnetic elements in the solar photosphere \citep{1995A&A...304L...1V, 2009Jess,2011ApJ...730L..37M, 2012ApJ...744L...5J, 2012ApJ...746..183J}; although oscillations and waves in faculae and network regions at lower spatial resolution have been studied more \citep{2008ApJ...676L..85K, 2009ApJ...692.1211C}.
\\
The small-scale magnetic features are expected to harbour a rich variety of waves excited largely by the buffeting they experience at the hands of the granulation and partly their interaction with the ubiquitous p-modes.
\cite{Hasan2003} argued that horizontal motion of magnetic elements in the photosphere can generate enough wave energy to heat the magnetized chromosphere. They found, based on numerical modelling, that granular buffeting excites kink waves and, through mode coupling, longitudinal waves. This was confirmed by \cite{Musielak2003}. They in fact argued that although transverse tube waves do not generate observable Doppler signals, when observing vertical flux tubes at disk centre, they excite forced and free longitudinal oscillations through a non-linear coupling. These should be observable if data with sufficiently high spatial resolution and sensitivity are available.\\
From the observational point-of-view, \citet{1995A&A...304L...1V}, using high spatial and temporal resolution spectropolarimetric data, detected short-period longitudinal waves ($P \simeq 100$ s) in small magnetic elements in the solar photosphere and estimated the energy flux they carried to be sufficient for the heating of the bright structures observed in the chromospheric network. \citet{2011ApJ...730L..37M} have found, using SUNRISE/IMaX data, magnetic flux density oscillations in internetwork magnetic elements, which they interpreted to be due to granular forcing.\\ 
In this work, we study the longitudinal and transversal oscillations of small magnetic elements in the solar photosphere and their power spectra.
To do this, we exploit the unprecedented combination of high spatial and temporal stability provided by SUNRISE/IMaX, together with the large number of observed magnetic features, to infer statistically significant information on oscillations of small-scale flux-tubes and the interaction between longitudinal waves and transverse kink waves. A particular advantage of this approach is the proven ability of SUNRISE to resolve small magnetic elements \citep{2010ApJ...723L.164L}, and to detect some of the internal structure of network features \citep{2012ApJ...758L..40M}. \\
This paper is organized as follows. In section $2$ we outline the data set used in this work, as well as the MHD simulation used to check the validity of our observational results. In section $3$, we describe the method used for the analysis of the oscillations which is based upon FFT analysis and wavelet analysis. In section $4$, we present the results. Starting from the FFT power spectra obtained from a limited number of magnetic features, we then apply wavelet analysis to a single magnetic element as a case study and then extend this analysis to the full sample of magnetic features at our disposal, to obtain statistically significant phase lag information between the longitudinal and transversal velocity. Section $5$ is devoted to the discussion of the results, while our conclusions are given in section $6$.

\section{Data sets employed: SUNRISE/IMaX observations and MHD simulations}
\begin{figure*}[]
   \centering
   \subfigure[IMaX continuum intensity] {\includegraphics[width=6cm,trim=12mm 3mm 12mm 3mm, clip]{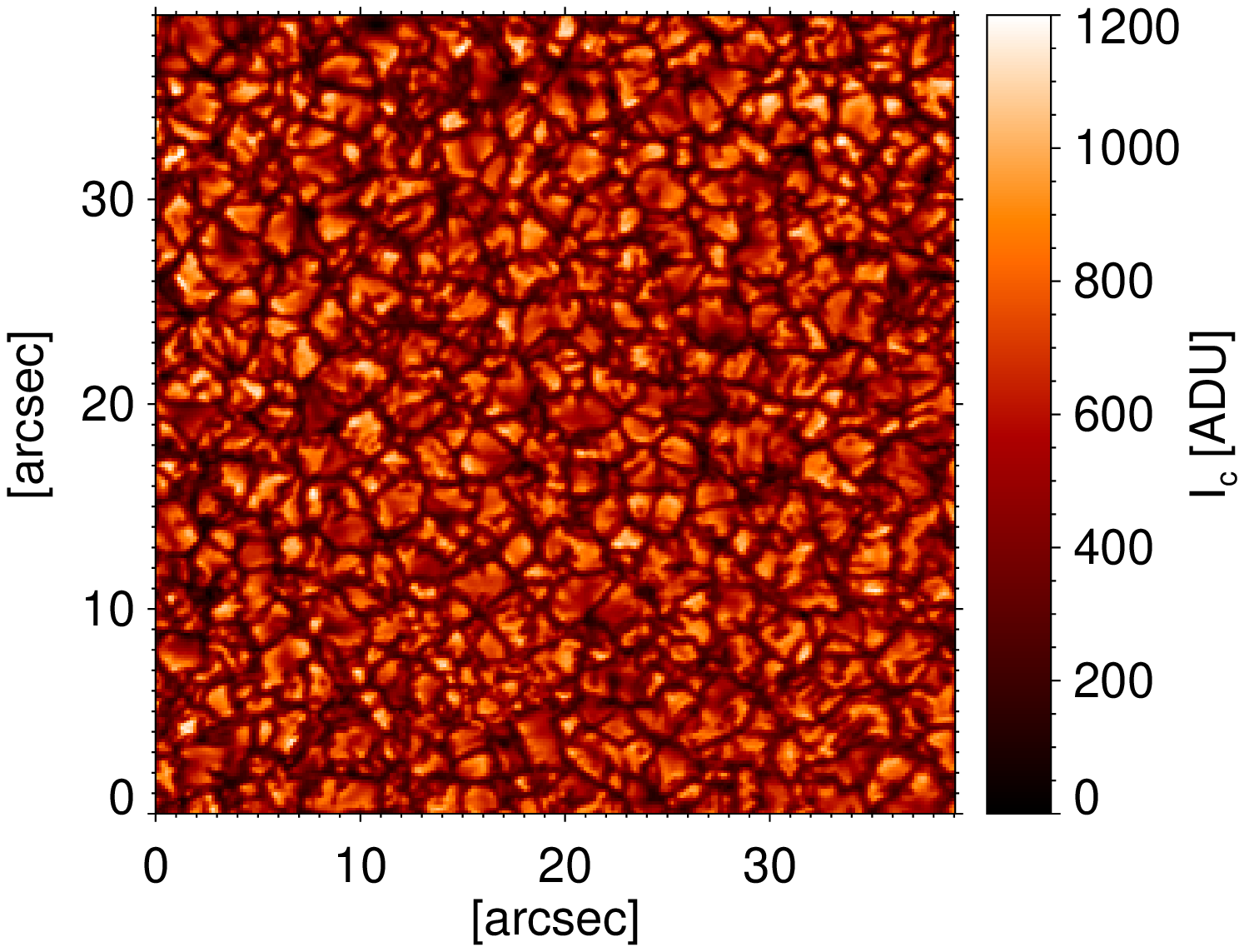}}
   \subfigure[IMaX longitudinal field]{\includegraphics[width=6cm,trim=12mm 3mm 12mm 3mm, clip]{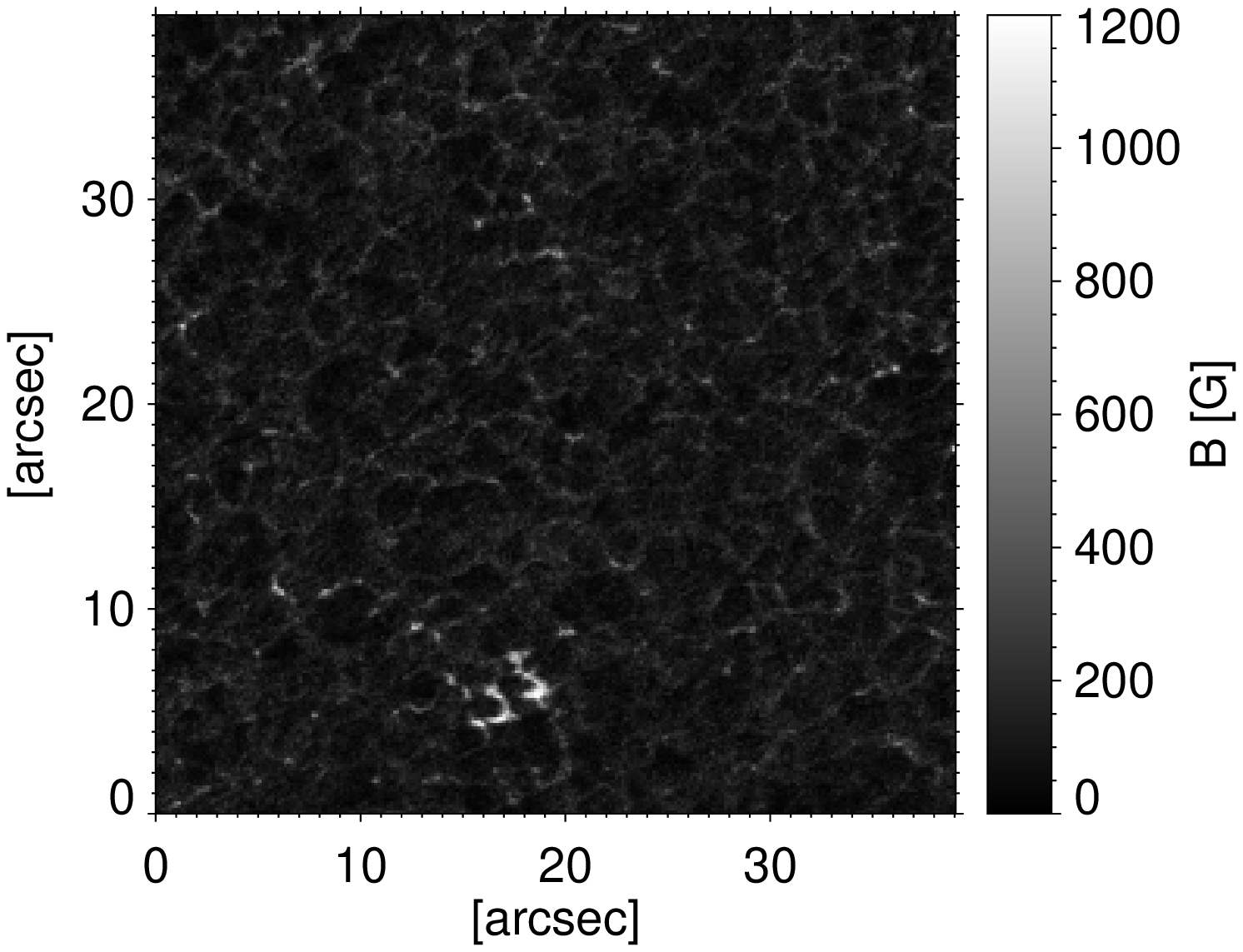}}
   \subfigure[$B_{z}$ MHD simulation] {\includegraphics[width=6cm,trim=12mm 3mm 12mm 3mm, clip]{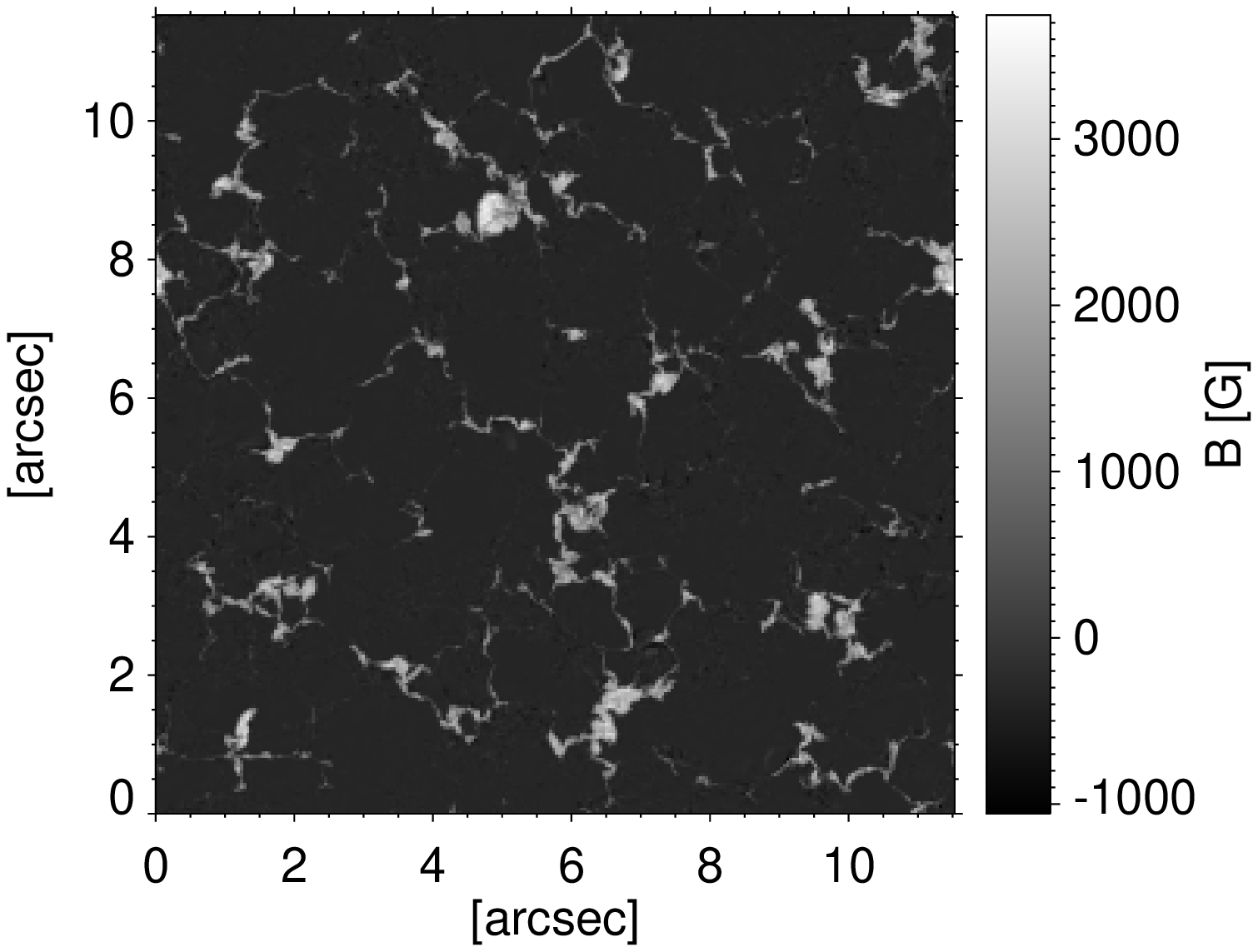}}
   \subfigure[IMaX features] {\includegraphics[width=6cm,trim=12mm 3mm 12mm 3mm, clip]{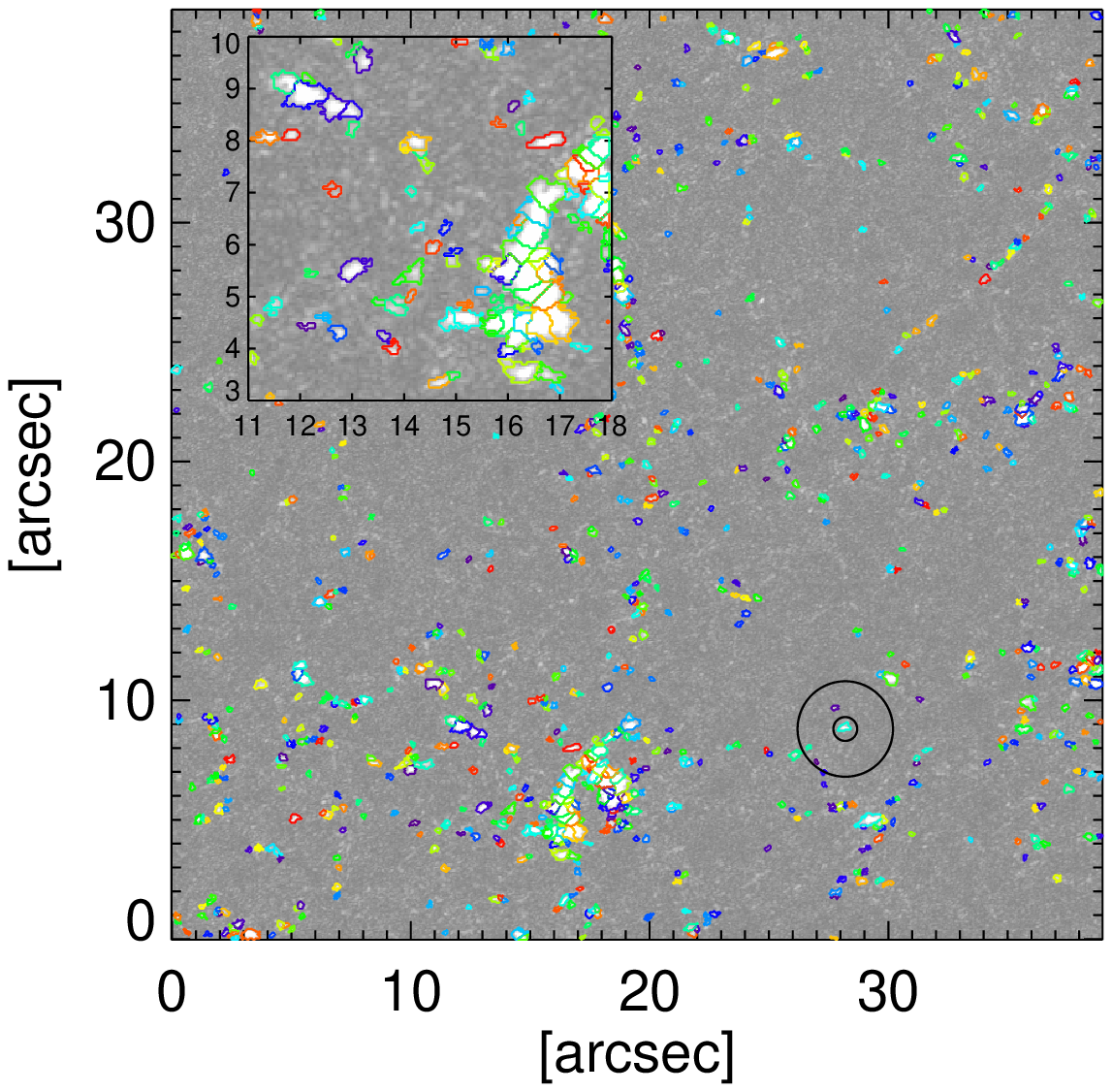}}   
   \caption{(a) IMaX continuum image. (b) IMaX longitudinal field obtained from SIR inversions. (c) MHD simulation: longitudinal component of the magnetic field. (d) IMaX features tracking example. The region between the two concentric circles, in the lower right part of the image, illustrates the size of the aperture placed around each magnetic element to determine the properties of the p-mode oscillations near that magnetic element.  Each coloured contour represents a label given to each identified magnetic feature (see main text). The inset displays a blow-up of a part of the field of view, including the left part of the large network patch near the bottom of the full image.} 
    \label{maps}
   \end{figure*}

The data set used in this work consists of a 2D spectropolarimetric time series with a length of approximately $32$ min and a cadence of $33$ s, acquired by the Imaging Magnetograph eXperiment (IMaX; \citealt{2011SoPh..268...57M}) on board the SUNRISE balloon-borne mission \citep{Barthol2010} in the Fe I $525.02$ nm spectral line. The data were taken on 2009 June 9 and encompass a quiet Sun region of approximately $40 \times 40$ arcsec close to disk center. In addition to common calibrations, the data were phase-diversity reconstructed \citep{2010ApJ...723L.127S, 2011SoPh..268...57M}, resulting in a spatial resolution of $0.15-0.18$ arcsec. The calibration procedure of IMaX data is summarized by \cite{2010ApJ...723L.175R}, although we used spectropolarimetric inversions instead of a gaussian fit to estimate the Doppler velocity in small magnetic elements. There is an intrinsic advantage in doing this. In magnetic elements, the gaussian fit can often fail as the Stokes-I signal is reduced. The inversions, however, also include Stokes-V profiles, which are in turn very strong. The zero crossing wavelength point of Stokes-V is a velocity indicator available to the inversion code but not to the gaussian fits.  For more details on the spectropolarimetric inversions we refer to  \citet{0004-637X-745-2-160}.
In Fig.~\ref{maps} we plot the continuum intensity map (panel a) and the longitudinal magnetic field (panel b) estimated by means of SIR spectropolarimetric inversions \citep{1992ApJ...398..375R}, assuming a single-component atmosphere with height-independent magnetic vector. \\ 
The MHD simulations analysed here were carried out using the MURaM code which solves the compressible MHD equations with an energy equation including radiation transfer in a non-grey approximation, and an equation of state which includes the effects of partial ionization \citep[see][for a full description of the equations and the numerical details]{2005A&A...429..335V}. The code has been used in for both quiet-Sun studies \citep[e.g.][]{2004ApJ...607L..59K,2007A&A...465L..43V} as well in studies of magnetic structures (for example pores, \citet{2007A&A...474..261C}, sunspots \citet{2009Sci...325..171R}, and active regions \citet{2007A&A...467..703C} amongst others). Here we use a simulation domain which is $24$ Mm in both horizontal directions and $1.4$ Mm in the vertical direction. The grid spacing is $20.8$ km in both horizontal directions, and $14$ km in the vertical. The top boundary is open to flows, and the magnetic field there is forced to be vertical. In this study we only use the vertical magnetic field and vertical velocity from the $\tau_{Ross}=1$ surface (at a height of about $800$ km above the bottom of the box), which was stored every $5$ to $7$ seconds. The total duration is $58$ minutes in total.

\begin{figure}[!t]
\centering
{\includegraphics[width=7cm]{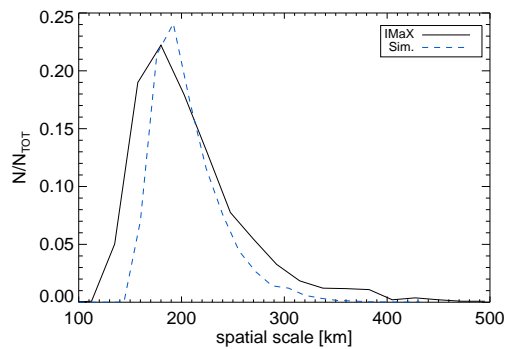}}   
\caption{Histograms of size of the analysed magnetic elements in the IMaX data (continuous line) and the simulation (dashed line). The size of each magnetic element is defined as the equivalent diameter associated to a circle with the same area as the measured one.}
\label{sizes}
\end{figure}

\section{Methods}
\subsection{Features tracking}
With the aim of studying the properties of the oscillations in small magnetic structures, we tracked the magnetic features in both the simulation and the IMaX data using the YAFTA code (Yet Another Feature Tracking Algorithm); a labelling flux-ranked uphill gradient algorithm \citep{Welsch2003, 2007ApJ...666..576D}. The algorithm tracks and labels groups of pixels lying on the same 'hill' in circular polarization maps, in the case of IMaX data, and vertical magnetic field maps in the case of simulations. For this purpose we set up two thresholds to avoid spurious detections in the final results. The first detection threshold is set at $2\sigma$ to both Stokes $V$ maps and $B_{z}$ maps, where for the observations the $\sigma$ value corresponds to the noise at the IMaX continuum wavelength, whereas for the simulations it corresponds to the standard deviation of the magnetic signal over the FoV. It is worth to mention here that the choice of $\sigma$ may affect the final results of our investigation. For this reason we also did a sensitivity test using different thresholds (see Sect. $4.3$). The second threshold acts spatially, allowing the detection of only those magnetic features whose area is larger than $9$ pixels (slightly larger than the spatial resolution achieved) in the case of IMaX data, and $49$ pixels for the MHD simulation. These two values correspond to a linear spatial scale of $~150$ km and $~140$ km respectively. This thresholds have been chosen in such a way that the distributions of size of the magnetic elements, in the simulation and in the IMaX data, are comparable (see Fig. \ref{sizes}). The size of each magnetic element is computed as the equivalent diameter of the circle with the same area as the measured one. The position of each tracked magnetic element is obtained from the center of gravity, and from that the velocity is thus estimated. \\
As extensively discussed in \citet{2013A&A...549A.116J}, the quantity $|B cos (\gamma)|$, with $\gamma$ being the magnetic field inclination, is retrieved more reliably than $B$ alone, since the latter can be affected by the noise in Stokes $Q$ and $U$. For this reason the use of Stokes $V$ maps in the case of IMaX observations is preferred.
We are aware, however, that using $B_z$ for the simulations and Stokes $V$ for the observations does introduce a certain bias. Thus, we expect that we choose fewer strong-field features in the observations, since these generally show a significant line-weakening due to the high temperature \citep[e.g.][]{1986A&A...168..311S,2010ApJ...723L.164L}.\\ 
The visual inspection of the results of the tracking revealed a number of short-lived magnetic features which were observable (traceable) for only a few time steps. Therefore, an additional temporal threshold was added to exclude them from the analysis, since we are mainly interested in improving the frequency resolution in the spectral domain. All the features lasting for less than $150$ s in both the simulation and the observational data set were rejected. 
In Fig. \ref{maps} (panel d) we mark the features fulfilling the criteria in a SUNRISE/IMaX snapshot by coloured contours. Each color represents a label given to the identified features.\\
The inset highlights that the algorithm tends to break up larger magnetic features into smaller ones, according to the number of maxima of the signal present there.
   \begin{figure*}[]
\centering
\subfigure[Horizontal velocity ]{\includegraphics[width=7cm]{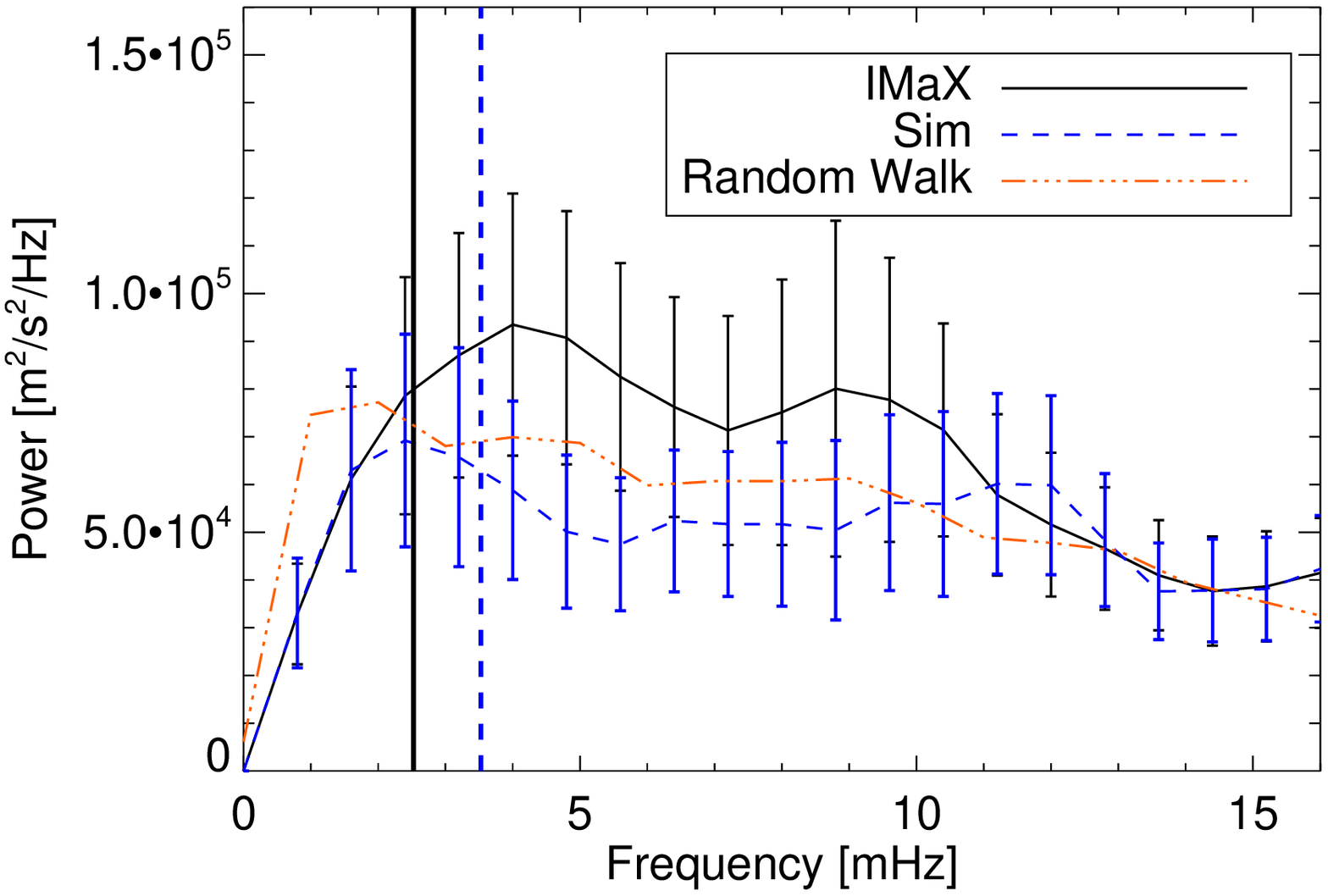}} 
\subfigure[Vertical velocity]{\includegraphics[width=7cm]{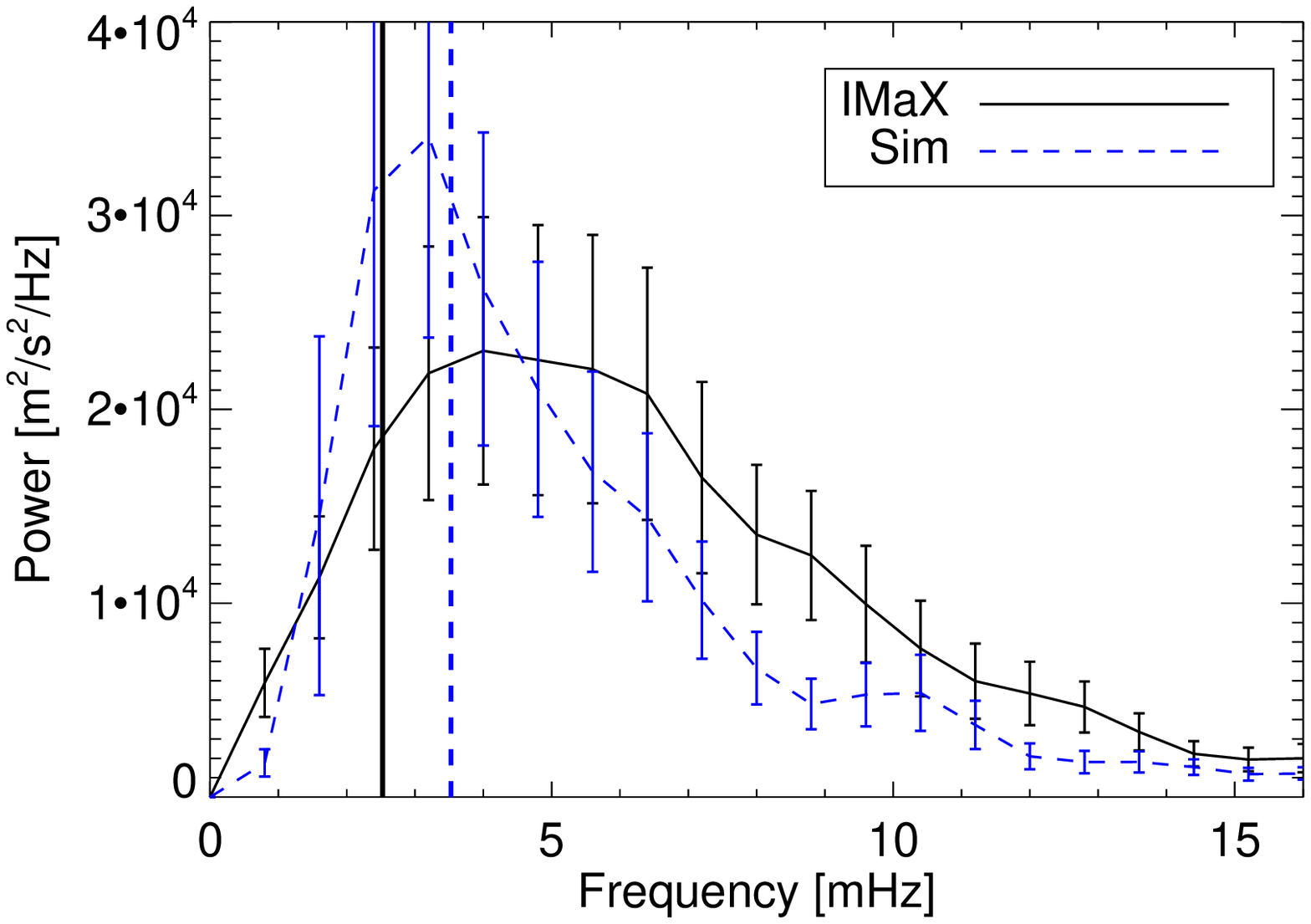}}   
\subfigure[Amplitudes horizontal velocity]{\includegraphics[width=7cm]{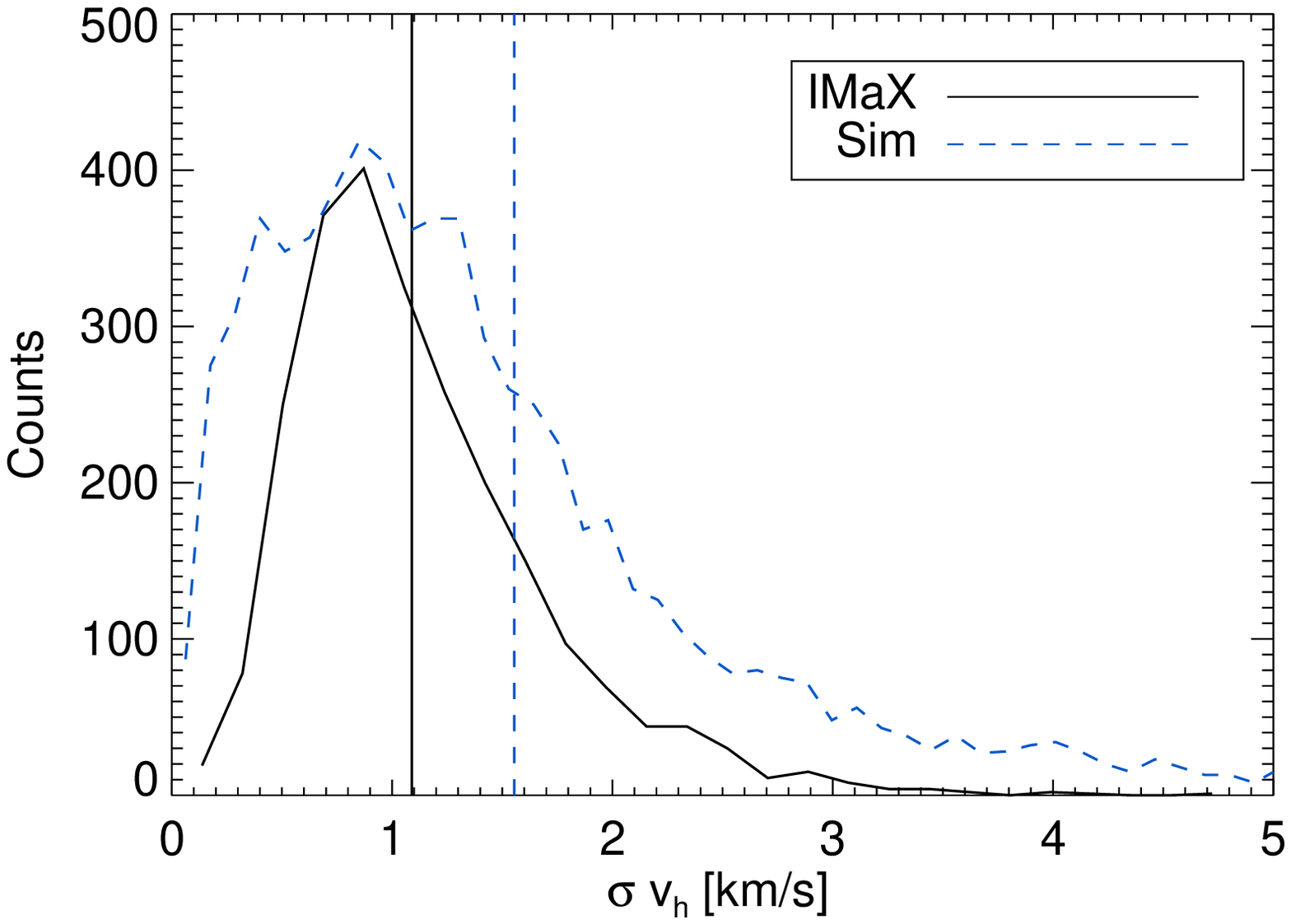}} 
\subfigure[Amplitudes vertical velocity]{\includegraphics[width=7cm]{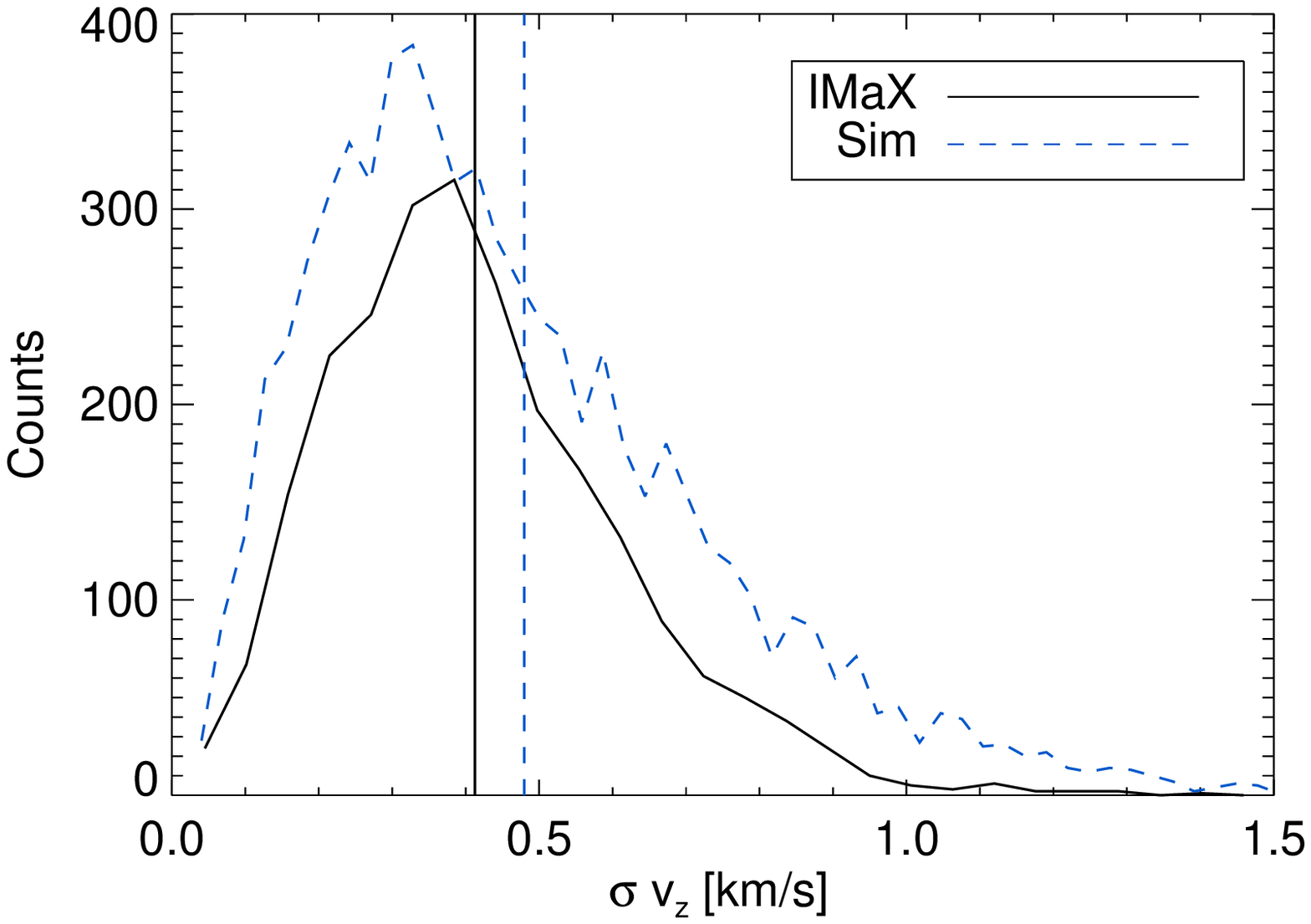}}   
\caption{(a) Power spectra of $v_{h}$ obtained from IMaX (solid line) and simulations (dashed line). The dot-dashed line represents the average power spectrum of $500$ simulated random walks. (b) Power spectra of $v_{z}$ obtained from IMaX and simulations. Each spectrum is the average of spectra of twenty individual features.  The vertical continuous and dashed lines indicate the lowest frequency due to the total length of the time series for the vertical velocity and the horizontal velocity, respectively. (c) Amplitudes of $v_{h}$. (d) Amplitudes of $v_{z}$. The vertical continuous and dashed lines indicate the mean value for IMaX and the simulation respectively. } 
\label{histograms_wave_amplitudes}
\end{figure*}
\subsection{Analysis of the oscillations}
For each magnetic element tracked by the algorithm, we estimated the vertical velocity ($v_{z}$) within its area, by averaging over values from each individual pixel, and the horizontal velocity ($v_{h}$) obtained by following its position. The first one corresponds closely to LoS velocity in the case of IMaX data, since the FoV is very close to disk center. The Doppler velocity was taken from the SIR inversions. In addition, we also estimated the contribution of the local non-magnetic oscillatory field ($v_{LNM}$) by considering the surroundings of each magnetic element, to further check the results. To this end, we surrounded each magnetic element with an aperture like the one shown in Fig. \ref{maps} (lower right part of panel d) between the concentric circles with inner and outer radii of $380$ km and $950$ km, respectively. The ambient oscillatory signal  $v_{LNM}$ is estimated by taking the average of the velocity of the non-magnetic pixels (below $10$ G) within the defined aperture. This prevents the contamination by possible magnetic field effects.\\
Our analysis consists of three steps. At first, we selected suitable magnetic features, on the basis of their lifetimes, to estimate the average power spectral density for both longitudinal and transversal velocity oscillations. We chose twenty cases for both simulations and observations with similar long lifetimes. We restricted ourselves to a few examples, picking particularly long-lived magnetic features, that can harbour also lower frequency waves, unlike the short-lived  majority of the magnetic features.  For each magnetic feature we estimated the periodogram using the Welch method \citep{blackman1958measurement}. Each periodogram is then interpolated over the same frequency grid, to take into account for the different lengths of the time series used, although the lengths of the time series chosen was very similar (on average $\sim 12-13$ min). We then estimated the average power spectral density for both longitudinal and transversal oscillations. In this case the simulation was resampled to the IMaX cadence ($33$ s).\\
In the second step of our analysis, we took the longest lived magnetic feature, among the twenty chosen before, as a case study. The amplitude of the oscillations of $v_{z}$ and $v_{h}$, as well as their coherence and phase were studied using wavelets. 
Since the time series associated to the transverse and longitudinal velocity are expected to be non-stationary, the wavelet analysis presents many benefits over the Fourier-based analysis, in particular, with respect to the estimate of the phase \citep{2004ApJ...617..623B}. Our analysis was performed with the standard tool by \citet{Torrence1999}. We used the complex Morlet mother function, which is also suitable for estimating the phase between two signals. In addition, we have also made use of the coherence spectrum. While the wavelet spectrum gives information about the distribution of power in the frequency-time domain, the coherence of two time series, being dependent on the cross-wavelet transform, is useful for investigating the interaction between the two physical mechanisms \citep{Torrence1999}; in our case between $v_{h}$ and $v_{z}$.\\
   \begin{figure*}[t]
   \centering
  \subfigure[Simulation: wavelet $v_{z}$]{\includegraphics[width=6.cm]{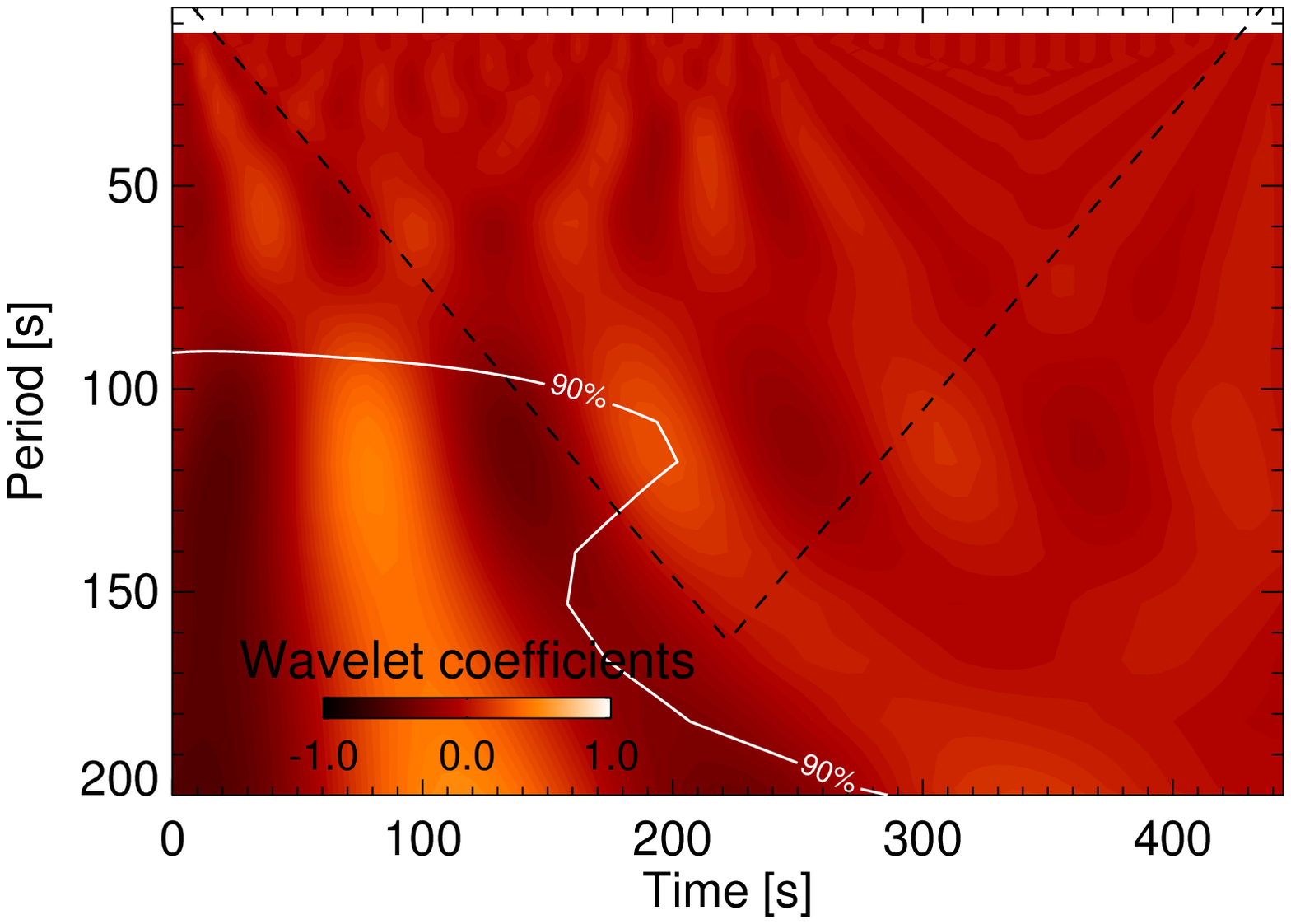}}
  \subfigure[Simulation: wavelet $v_{h}$] {\includegraphics[width=6.cm]{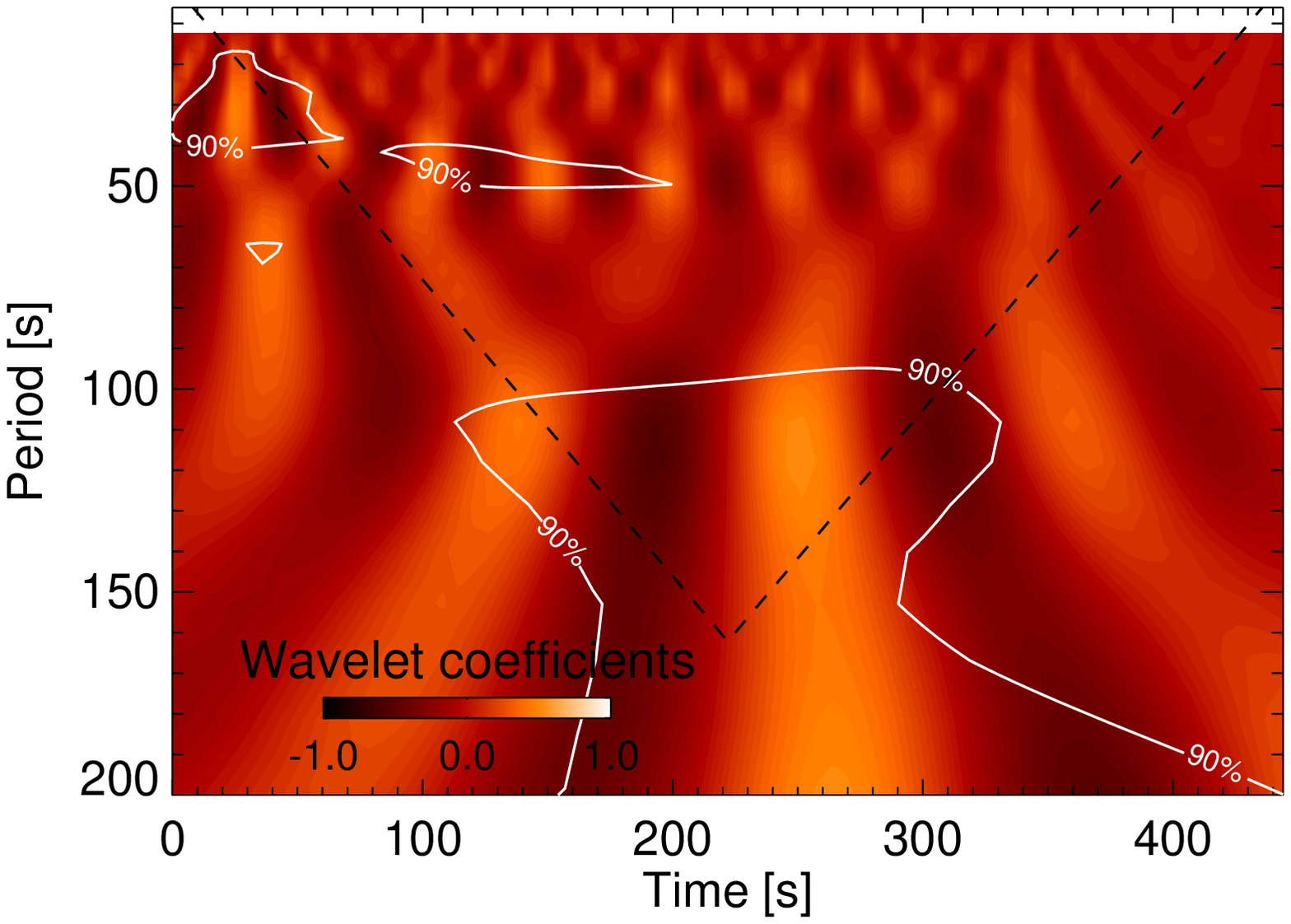}}
  \subfigure[Simulation: coherence $v_{z}-v_{h}$] {\includegraphics[width=6.cm]{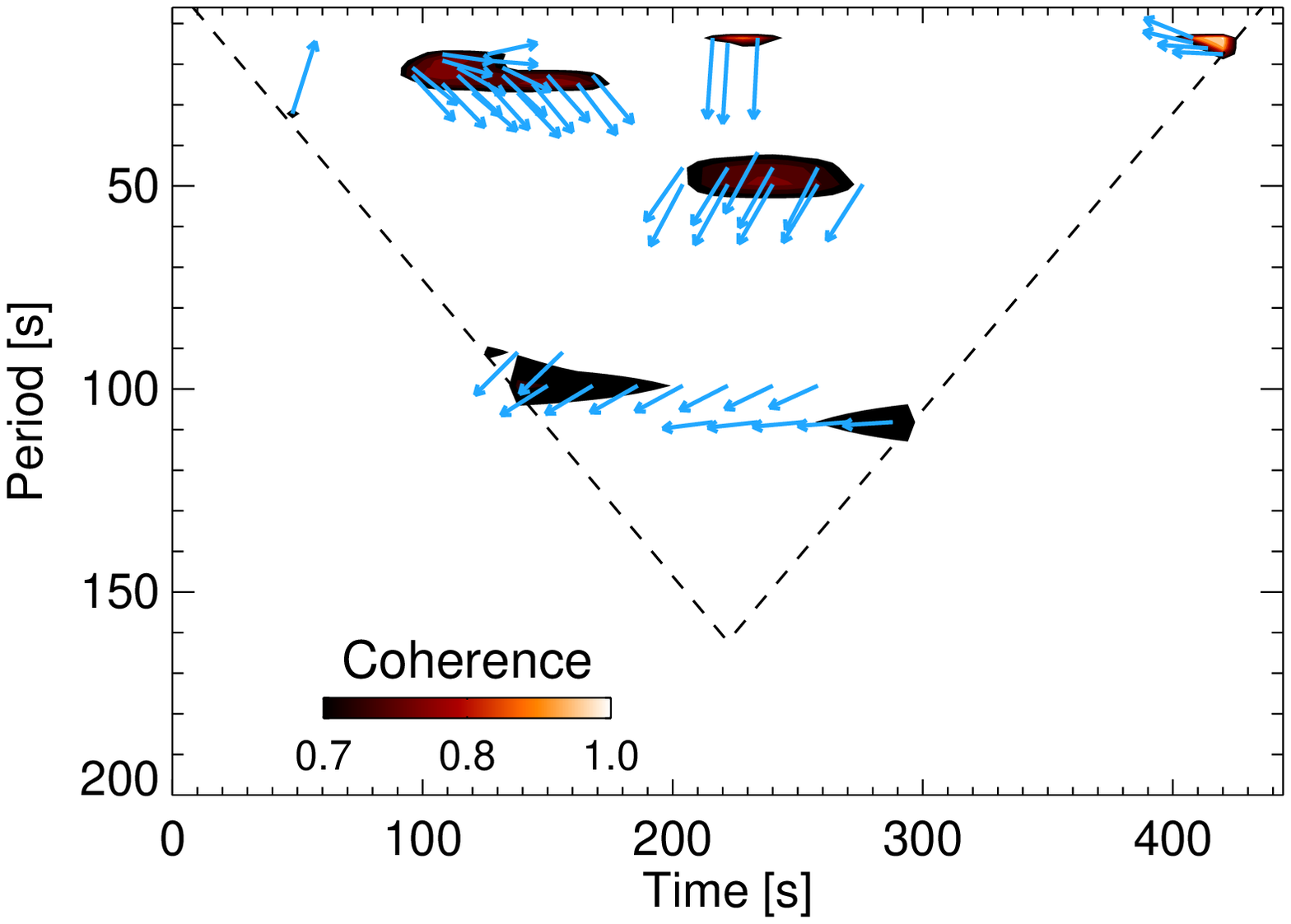}}
  \subfigure[IMaX: wavelet $v_{z}$] {\includegraphics[width=6.cm]{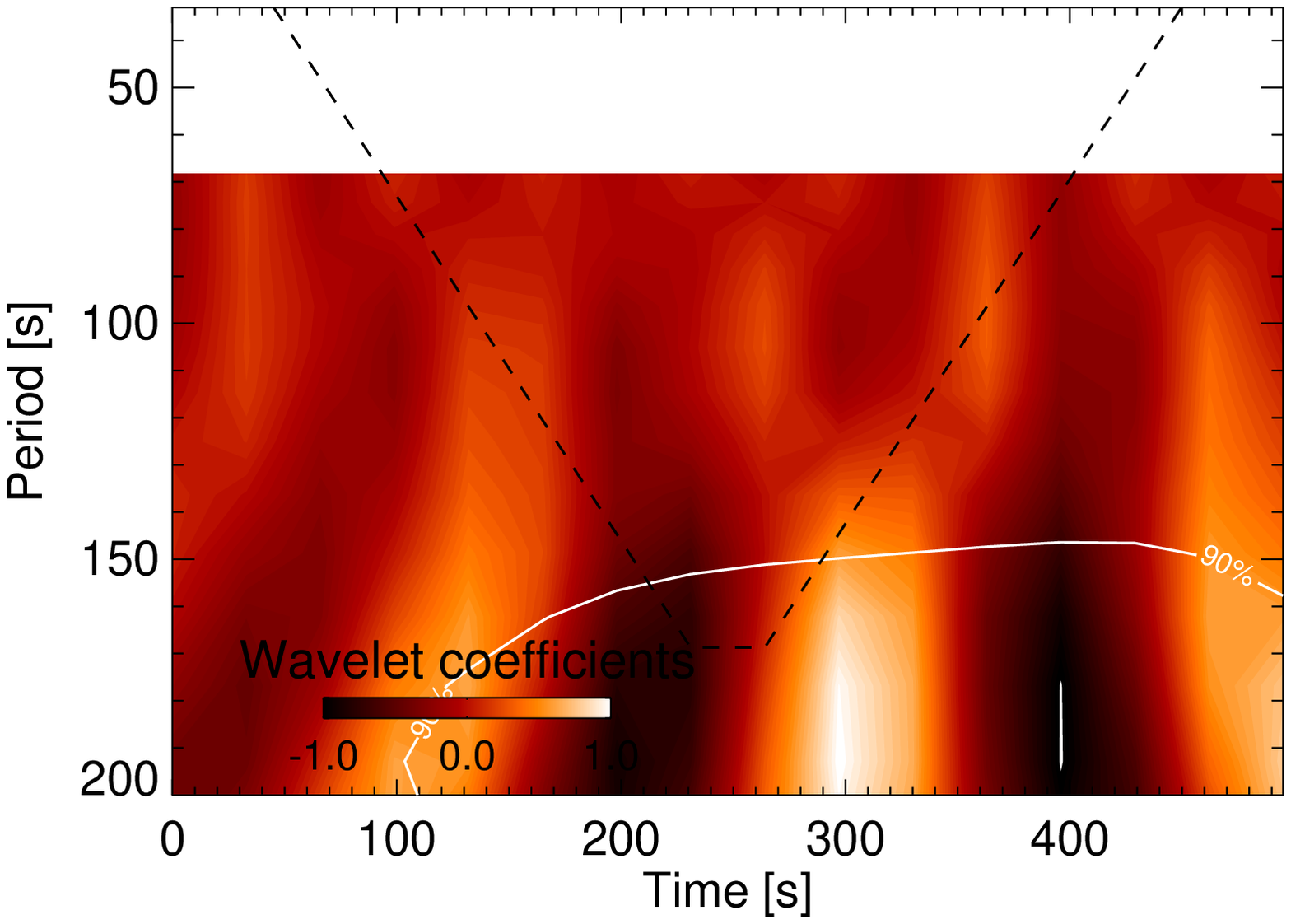}}   
  \subfigure[IMaX: wavelet $v_{h}$] {\includegraphics[width=6.cm]{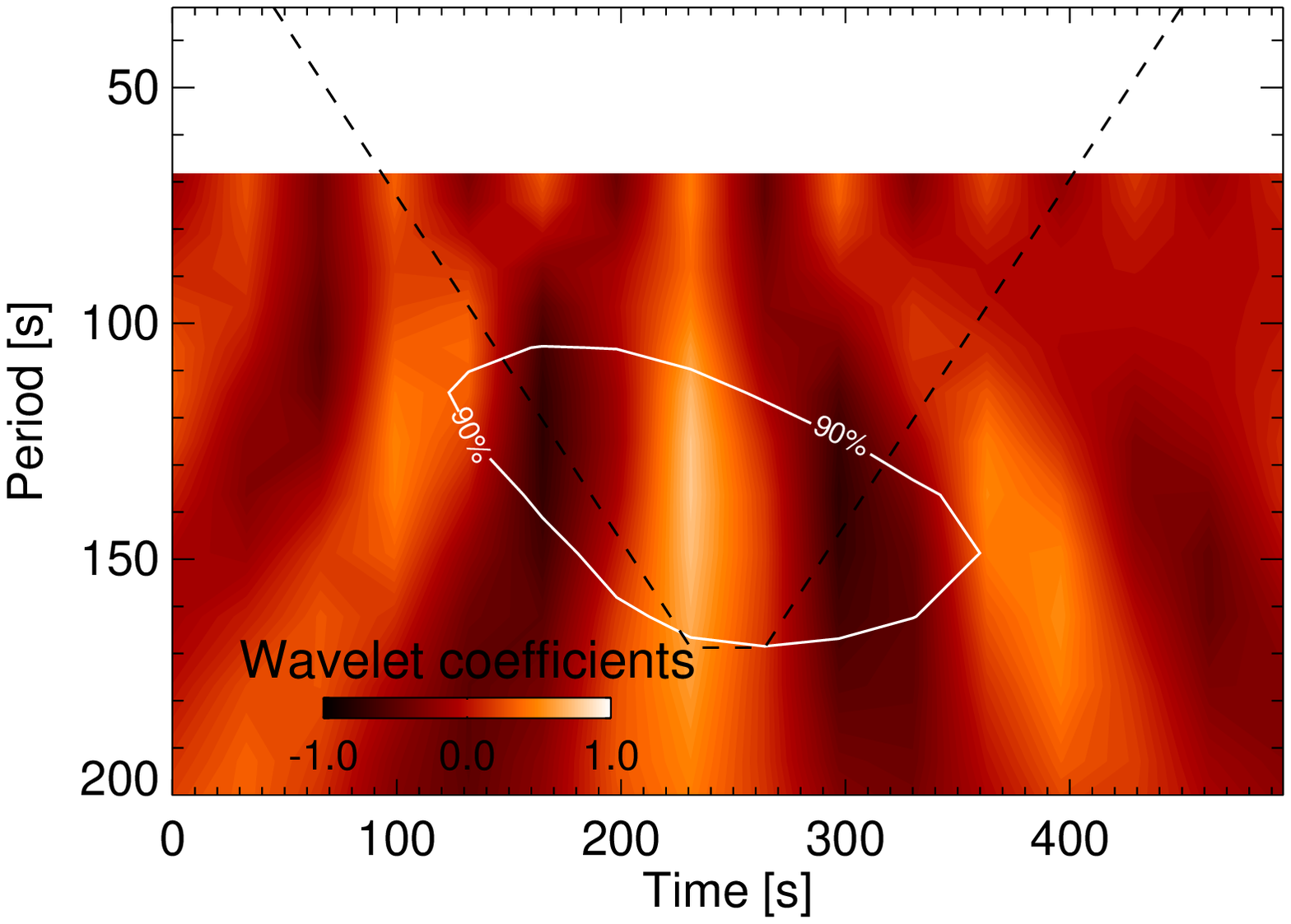}}   
  \subfigure[IMaX: coherence $v_{z}-v_{h}$] {\includegraphics[width=6.cm]{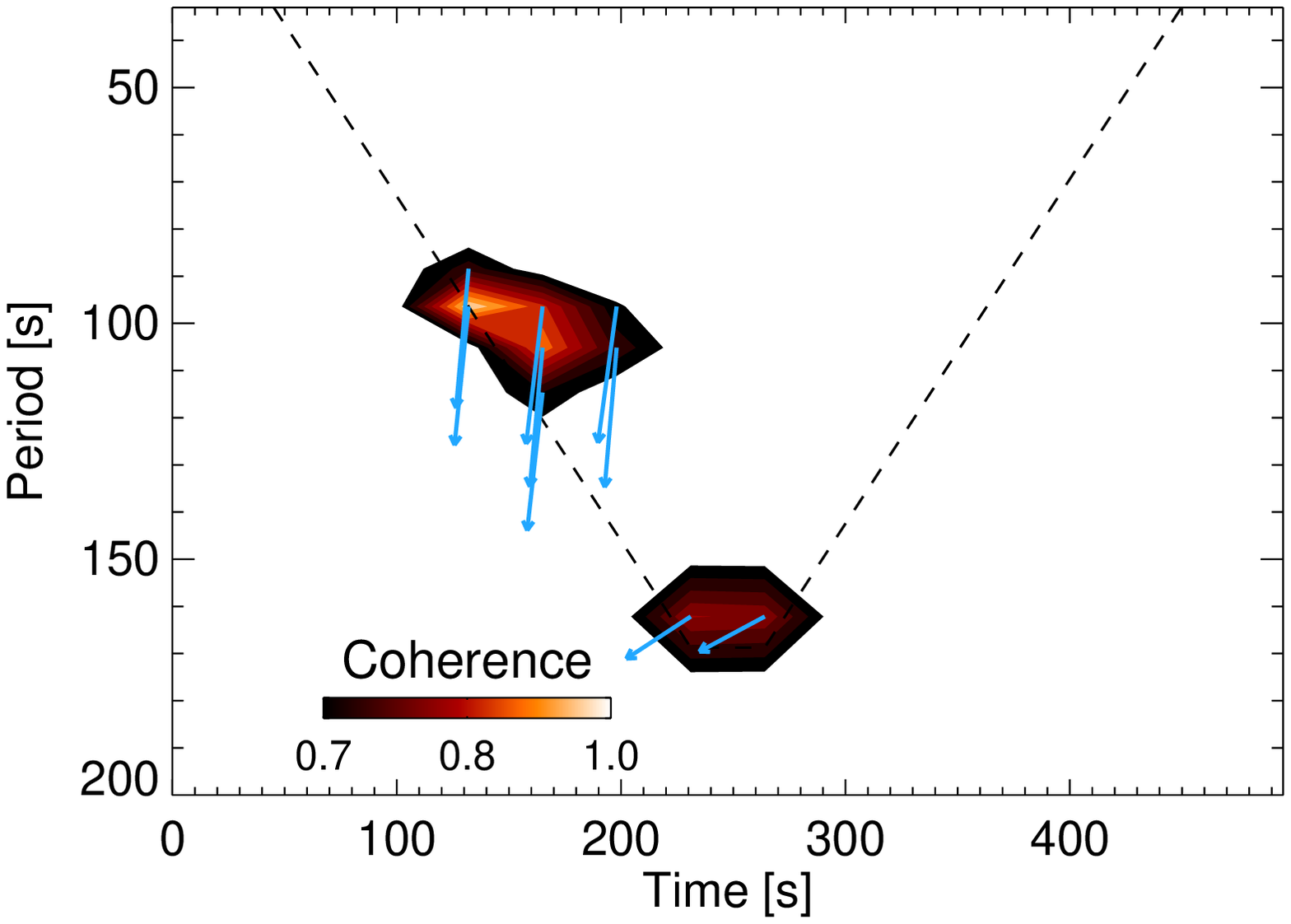}}
   \caption{(a) Wavelet amplitude spectrum of $v_{z}$ from the simulation. (b) Wavelet amplitude spectrum of $v_{h}$ from the simulation. (c) Wavelet phase coherence spectrum between $v_{z}$ and $v_{h}$ from the simulation. The arrows represent the phase lag in polar coordinates (zero phase corresponds to arrows pointing to the right). (d) Wavelet amplitude spectrum of $v_{z}$ from IMaX. (e) Wavelet amplitude spectrum of $v_{h}$ from IMaX. (f) Wavelet phase coherence spectrum between $v_{z}$ and $v_{h}$ from IMaX. The arrows represent the phase lag in polar coordinates (zero phase corresponds to arrows pointing to the right. The dashed line represents the cone-of-influence where results cannot be trusted (area between the dashed lines and the edges of the diagram). The white contours represent the $95\%$ confidence level.}
      \label{wavelet}
   \end{figure*}
The third and final step of our analysis represents an extension of the second step to the full population of magnetic features. In particular, we focused on the phase angle between $v_{h}$ and $v_{z}$.\\
The total number of features tracked in the IMaX data and the simulation amounts to $2384$ and $6388$, respectively. The larger amount of features tracked in the simulation can be explained in terms of a combination of effects. Thus, the estimated average lifetime of the magnetic elements in the simulation is $2.5-3$ times smaller than the average lifetime of the magnetic features collected in the IMaX data. This fact, together with the larger duration of the simulation ($58$ min) with respect to the IMaX data ($32$ min), and the intrinsic higher spatial  resolution of the simulation can significantly increase the number of features collected.  \\
Our goal is to obtain the distribution of the phase difference between the transversal and longitudinal velocity inside the magnetic elements. A clear phase relation would point to an interaction between horizontal and vertical velocity perturbations. A number of constraints, described below, were applied on the feature to ensure the reliability of the results. \\  
In the coherence diagram given by the wavelet analysis, we selected, for each magnetic feature, the location in the frequency-time domain at which the two time series under investigation show the highest coherence. The phase between the two signals is then estimated at this location. A phase estimate was considered only if the coherence between the two signals under investigation was above $0.8$ outside the cone-of-influence of the wavelet phase diagram, i.e. in the region where the wavelet analysis can be trusted. The temporal cadence of the simulations was reduced to match the IMaX cadence ($33$ s).\\
Besides determining the phase shift between the longitudinal velocity $v_{z}$ and the transversal velocity $v_{h}$, we repeated the analysis between the surrounding oscillatory field $v_{LNM}$ and $v_{h}$. The latter is done to check whether any phase relation between the longitudinal and transversal velocity is strictly inherent to the wave signal observed inside the magnetic elements, or not. \\
In addition, we also studied the distribution of periods at which $v_{z}$ and $v_{h}$ are coherent. This is done to find the frequency band at which their interaction (if any) takes place.

\section{Results}
\subsection{FFT power spectra and amplitude of oscillations}
Through the features tracking we obtained the horizontal and vertical velocity associated to each magnetic element.\\
As mentioned in Sect. $3.2$, we at first estimated the average power spectra of the twenty magnetic features characterized with the longest lifetimes. \\
In panels (a) and (b) of Fig. \ref{histograms_wave_amplitudes}, we show the power spectra obtained from the IMaX data and the simulation. The power spectrum for both horizontal velocity (panel a) and vertical velocity (panel b) are shown. The vertical line indicates, the poorer frequency resolution among the time series chosen (i.e. the frequency resolution of the shortest time series $\nu_{min}=2/T_{length}$). The region at low frequency below this line should therefore not be trusted.\\
Both simulation and observations display a similar behaviour. The power spectrum of $v_{z}$ shows that most of the power is concentrated at low frequency, in the $2-7$ mHz band. \\
The horizontal velocity, in turn, is characterized by a broader power spectrum with high-frequency components ($> 8-10$ mHz), although the highest power is still located in the low frequency band. The power of the horizontal velocity is larger than the power of $v_{z}$, at all frequencies, particularly in the IMaX data. Consequently the amplitude of the $v_{h}$ oscillations is larger. This can be seen in panels (c) and (d) of Fig. \ref{histograms_wave_amplitudes}, where we plot the distributions of the standard deviation of $v_{h}$ and $v_{z}$ obtained from the complete population of magnetic features. Both simulations and observations show an excellent agreement, with most of the magnetic features displaying horizontal oscillations having a mean amplitude around $1-1.5$ km/s, while for the longitudinal velocity, the corresponding values lie  in the range $0.4-0.5$ km/s. 

   \begin{figure}[]
\centering
\subfigure{\includegraphics[width=6cm]{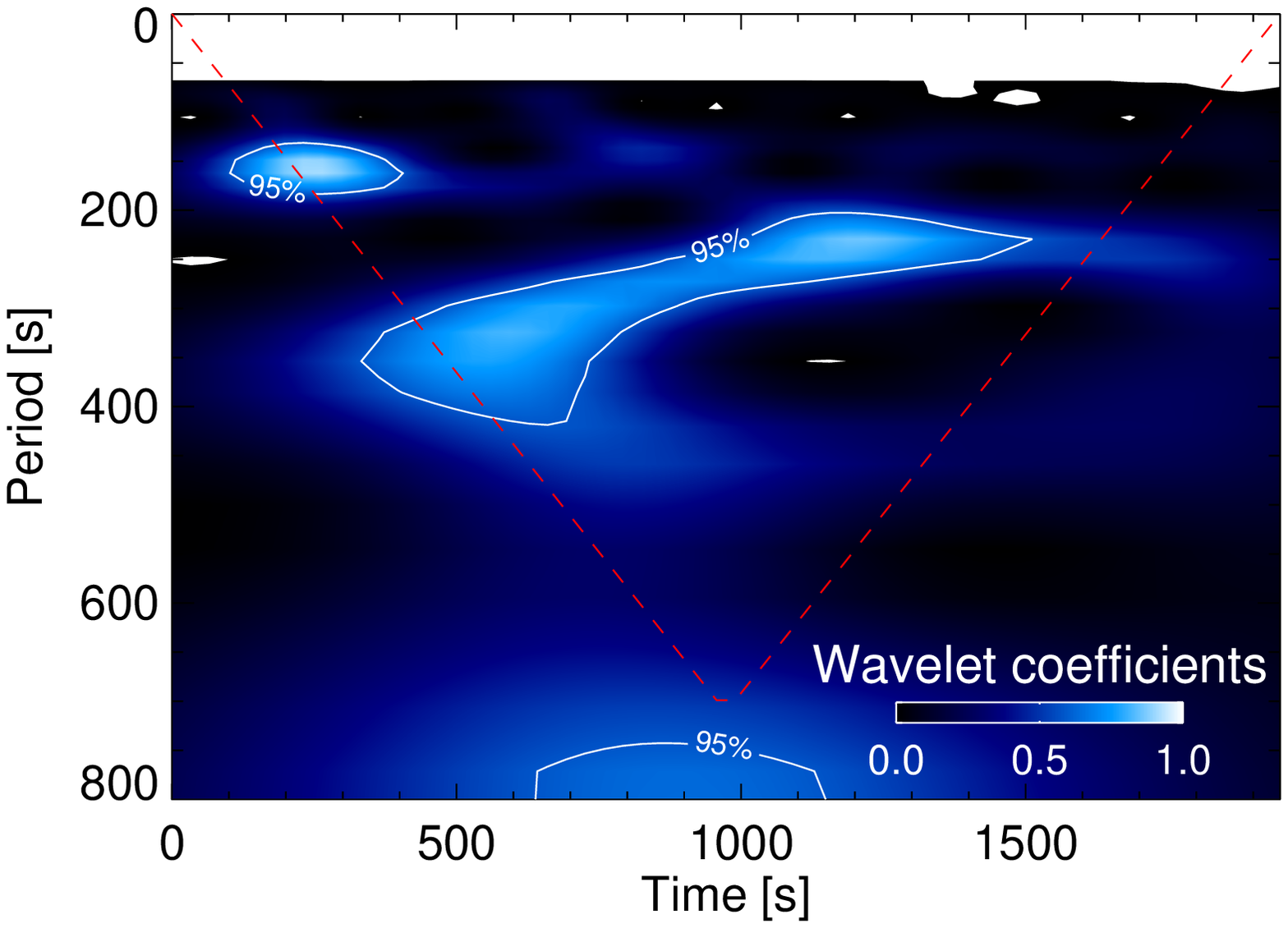}} 
\subfigure{\includegraphics[width=6cm]{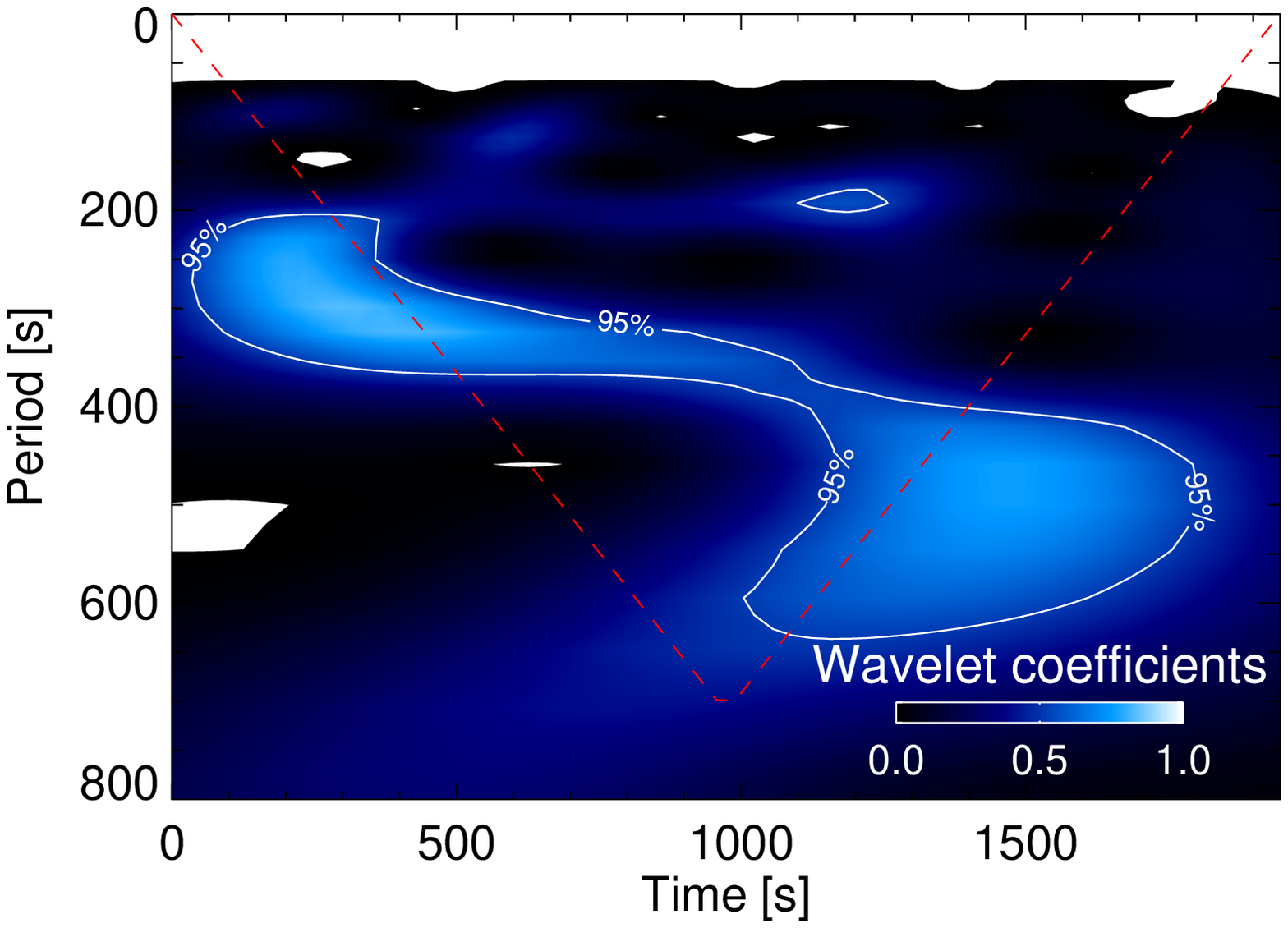}}   
\subfigure{\includegraphics[width=6cm]{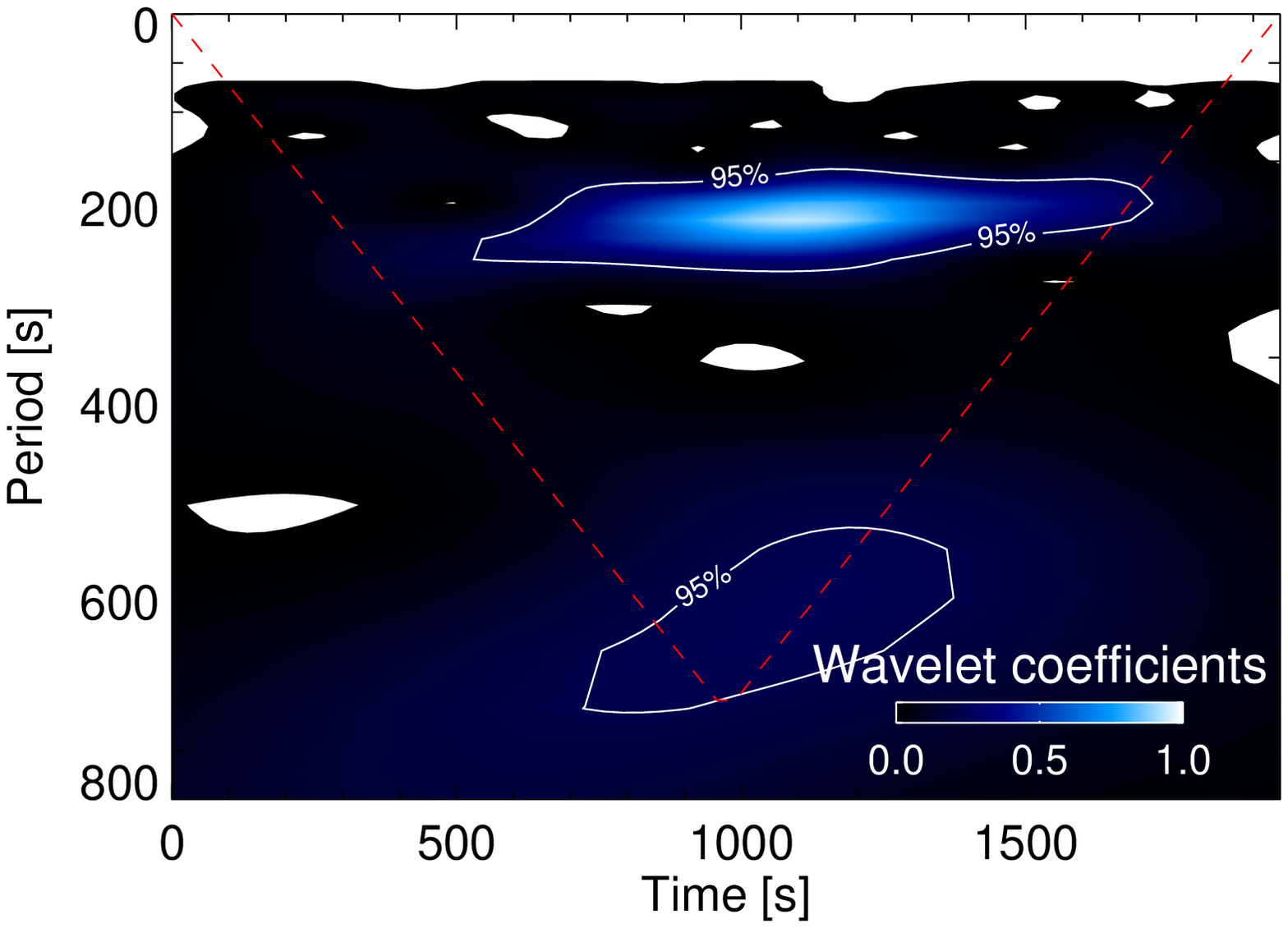}}    
\caption{Power spectra of three different simulated random walks. The continuous line represents the  $95\%$ confidence level, the dashed line represents the cone of influence. The Nyquist period is set by the sampling to $66$ s.} 
\label{randomwalks}
\end{figure}

\subsection{Wavelet amplitude analysis and coherence: a case study}
\label{wavelet_case}
Fig. \ref{wavelet} depicts the wavelet diagrams for the two longest time series found respectively in the IMaX data and in the simulation domain (i.e. the longest lived magnetic element in each).
The time series associated with the magnetic element found in the IMaX data has been truncated to match the lifetime of the magnetic element from the simulation. This is to force the wavelet diagrams to be comparable in terms of frequency resolution.\\
Panels (a) and (d) show the amplitude spectrum of $v_{z}$, for the simulation and the observations, respectively. 
Panels (b) and (e) show the amplitude spectrum of $v_{h}$ and, panels (c) and (f), the coherence between $v_{z}$ and $v_{h}$. The arrows represent the phase angle between the two signals in polar coordinates, where zero is along the $x-axis$. In our sign convention, a negative phase angle (arrows pointing down) means that $v_{h}$ leads $v_{z}$. In the same plots, the white contour represents the $90\%$ confidence level, while the dashed line represents the cone of influence.\\ 
The vertical velocity, $v_{z}$, is characterized in both the simulation and the IMaX data by oscillations with periods larger than $130-150$ s. It is worth noting that, although the fiducial line extends to longer periods in the three-minute band, only the region outside the cone-of-influence can be trusted and, therefore, only this region will be considered in our analysis.\\
The horizontal velocity $v_{h}$ of the magnetic elements show power in the band $P=100-150$ s and at higher frequency. In particular in the simulation, where the Nyquist frequency is larger, one finds many peaks above the confidence level close to $P=50$ s or even smaller periods. This result is compliant with what has found in the power spectral density (Sect. 4.1).\\
In both the simulation and the IMaX data, one finds locations in the wavelet diagram where $v_{z}$ and $v_{h}$ are highly coherent (see panels (c) and (f) of Fig. \ref{wavelet}). As the coherence is a measure of the interaction between two signals, which depends on the cross-spectrum, one can find high coherence even outside the confidence levels of each time series. This means that even in the case of minimal power the processes associated to the vertical and horizontal velocity may still strongly interact.\\
A high coherence is found to reside at periods of $100$ s and shorter. The associated phase angle is always different from $0^{\circ}$. In the case of the simulation the phase lag is in the range  $(-70^{\circ},-110^{\circ}) $, while for IMaX data it is close to $-95^{\circ}$. \\

\subsection{Power spectrum of random walks}
Small flux tubes in the photosphere are subject to the forcing action of the granular flows. Their displacements and oscillations represent therefore the response to this forcing.\\ 
In this section we present the results of a simulation conducted to test whether the $v_{h}$ power spectra are compatible with random walks.\\ 
We simulated the paths of $500$ $2D$ random walks lasting for $2000$ s and with a temporal step of $1$ s. The physical size of the spatial grid was set to $0.5$ km. This is equivalent to set the amplitude of the fluctuation of the position in the time interval. This value was chosen a posteriori to ensure that the mean amplitude of $v_{h}$ of the simulated random walks was $\simeq 1.3$ km/s, thus similar to the real one at the same IMaX cadence (after resampling the simulated time series at $33$ s).
After setting up this value, we repeated the simulation. For each random walk we estimated the power spectrum of $v_{h}$. In Fig. \ref{histograms_wave_amplitudes} (panel a) we plot (dot-dashed line) the average power spectrum obtained from averaging over all the $500$ cases. The shape of the power spectra are similar, thus the observed oscillations of the flux tubes appear to be consistent with random walk displacements due to buffeting. For comparison with the real case discussed in the previous section, In Fig. \ref{randomwalks} we also show three examples of wavelet diagrams obtained from the simulated random walks. The power is distributed over a broad range of periods, from large periods at the frequency resolution imposed by the duration of the simulated time series, down to the Nyquist limit ($66$ s).

\subsection{Statistical analysis: histograms of phase}
\begin{figure}[!ht]
   \centering
  \subfigure[$\phi(v_{z}-v_{h})$] {\includegraphics[width=7.5cm]{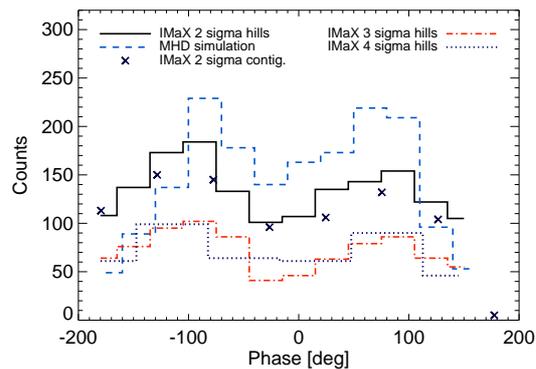}}
    \subfigure[Period distribution ] {\includegraphics[width=7.5cm]{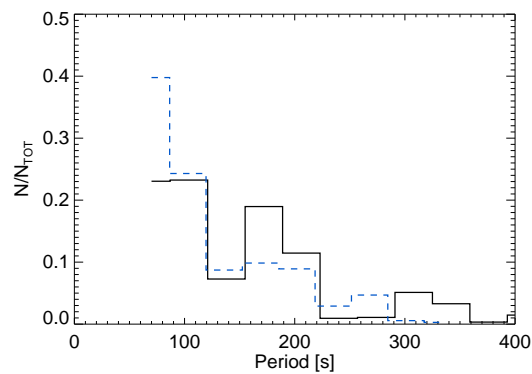}}

   \caption{(a) Histogram of phase $\phi (v_{z}-v_{h})$ obtained from IMaX with different thresholds ($2\sigma$ black continuous line, $3\sigma$ red dash-dot line and $4\sigma$ dotted line), and the simulation (dashed line). Crosses represent the histogram of phase obtained from IMaX data tracking of contiguous pixels instead of local maxima.
   (b) Normalized histogram of the periods at which the coherence, between $v_{z}$ and $v_{h}$, is largest for IMaX (continuous line) and the simulation (dashed line). } 
    \label{phase}
   \end{figure}

\begin{figure}[!t]
\centering
{\includegraphics[width=7.5cm]{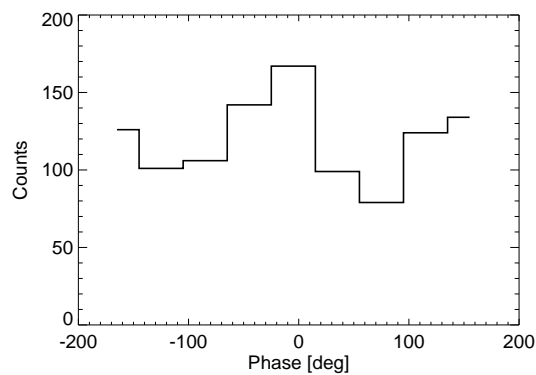}}   
\caption{Histogram of phase $\phi (v_{h}-v_{LNM})$ obtained from IMaX data.} 
\label{check}
\end{figure}

To find a statistically reliable estimate of the phase relation $\phi(v_{z}-v_{h})$, we used the full sample of tracked magnetic elements.\\
For this purpose, we used the wavelet analysis described in Sect. \ref{wavelet_case} to estimate the phase corresponding to the highest coherency in the wavelet diagram, for each tracked magnetic element. 
In Fig. \ref{phase}a we compare the results obtained from IMaX observations (continuous line) to those from the MHD simulation, resampled to match the IMaX cadence (dashed line). Both histograms display clear maxima at $-90^{\circ}$ and $+90^{\circ}$. \\
To test the consistency of this result, we studied the phase relation between the horizontal velocity and the longitudinal velocity surrounding the magnetic elements. The aim of this test is to check whether the $\pm 90^{\circ}$ phase angle is inherent to the velocity signal strictly inside the magnetic elements, or rather it is affected by either contaminations due to the surrounding acoustic field or some issue concerning the analysis. To this end we estimated the mean velocity signal in an annular region like that shown in Fig. \ref{maps}d, where we selected only the non-magnetic pixels. 
The results of this check are plotted in Fig. \ref{check}. The phase lag is zero in this case. This result demonstrates that the $\pm 90^{\circ}$ phase lag between the vertical and horizontal velocities is obtained only if the internal velocity field of the flux tube is considered. Therefore it is closely related to the nature of the MHD waves in the magnetic elements. As mentioned in the previous section, we tracked groups of pixels lying on the same 'hill'. This method tracks each magnetic element (peak in the Stokes $V$ signal) separately, even if it is part of a tight cluster (i.e. the magnetic signal stays high between individual elements. Clusters of magnetic elements may have a correlated horizontal velocity, hence their contribution to the histogram of phase may significantly affect its shape. To further test our results, we studied the histogram of phase obtained from the IMaX data but tracking contiguous pixels instead of local maxima. Thus, we considered all the contiguous pixels with Stokes $V$ above a given threshold to belong to a single magnetic feature. In this case, regardless the presence of local maxima, the magnetic structure is tracked as a whole. The resulting histogram is also shown in Fig. \ref{phase}a (crosses). Even in this case, the histogram of phase shows two maxima at $-90^{\circ}$ and $+90^{\circ}$.     \\
We also note that a significant fraction of elements show a small phase shift ($ \sim 23 \%$ in the IMaX case). This can be due to projection effects. If a magnetic feature is inclined to the LOS, there will be a component of the transverse velocity, associated to the observed kink oscillations, along the LOS which will be: $v_{z, projected}= v_{h} sin (\gamma)$, where $\gamma$ is the inclination angle. By the same token, the amplitude of the kink velocity will be decreased by $cos \gamma$.\\
If we consider the average observed horizontal velocity (see Fig. \ref{histograms_wave_amplitudes}), after the correction for the projection effects, we get $v_{z, projected} \sim 0.19-0.29$ km/s. This is estimated using an inclination angle between $10^{\circ}$ and $15^{\circ}$, as obtained by Jafarzadeh et al. (in prep.) from IMaX observations. \\
Form the histogram of the LOS velocity we can estimate the probability of finding a magnetic element with $v_{h}$ smaller than the above limits. This corresponds to the case in which the longitudinal velocity would be dominated by the projection of the horizontal velocity along the LOS, thus they would be in phase. We estimate this probability between $14\%$ and $29\%$, which is comparable with the observed fraction of elements with a small phase shift.
It has to be noted that our selection criteria based upon Stokes $V$ and $B_{z}$ in the IMaX data and the simulation, respectively, may result in a sample of magnetic features not restricted to vertical flux tubes, including a significant fraction of inclined features for which $\phi(v_{z}-v_{h})$ is close to zero because of the mechanism above. However, this may not be the only mechanism responsible for the large spread of the peaks in the histogram of phase. Another important cause may be the difference in the height of maximum response of Stokes $V$ to vertical and horizontal velocities in the magnetic elements. This mechanism may contribute, along with the first one cited above, to increase the number of elements with small phase lag.\\
 Moreover, as anticipated in Sect. $3.1$, we checked the results against possible effects due to the particular choice of the threshold used in the tracking code. In fig. \ref{phase} we also plot the results of this further analysis using a threshold of $2\sigma$, $3\sigma$ and $4\sigma$ on the IMaX data. As expected the histograms are shifted to lower values as the threshold increases, but their shape is not changed, showing again two maxima at $-90^{\circ}$ and $+90^{\circ}$.

We also studied the distribution of periods at which one finds the highest coherence between $v_{z}$ and $v_{h}$, that is periods at which the $\pm 90^{\circ}$ phase lag is found. We recall here that we only considered those cases in which the coherence was at least $0.8$. This is a demanding constraint. This distribution is shown in Fig. \ref{phase}b. The continuous line represents the IMaX data, the dashed line the simulation. It is evident that, in most of the cases, the interaction between the vertical and the horizontal velocity takes place  in the high-frequency band, at periods smaller than $200$ s.\\

\section{Discussion}
\label{discussions}
The presented results concern the main properties of velocity perturbations in small magnetic elements in the solar photosphere.
In particular we have studied transverse oscillations and longitudinal velocity oscillations of these magnetic elements. Starting from the longest time series associated with different magnetic features, we have retrieved the average power spectral density for the two kinds of oscillations. Since most of the magnetic features tracked, and their associated time series, have a short lifetime (shorter than $10$ minutes), the wavelet analysis constitutes a more suitable tool for the analysis of the periodicities and, more specifically, for the estimate of the phase angle between $v_{z}$ and $v_{h}$. This is even more important when dealing with non-stationary signals like those expected for these kinds of velocity oscillations.\\ 
Our results demonstrate that, while $v_{z}$ is characterized by comparatively low frequency oscillations ($\nu < 7-8$ mHz, i.e. periods larger than $120-140$ s) and amplitudes of the order of $500$ m/s, transverse displacements have a more extended power spectrum with very high frequency peaks ($\nu \sim 8-10$ mHz, i.e. periods of $120-100$ s), and with amplitudes of the order of $1$ km/s. Both these modes are expected to be propagating at these frequencies with significant observed power.\\
The wavelet analysis has also revealed that the horizontal velocity is characterized by high frequency wave packets, confirming the non-stationarity of the signals. Unfortunately, the short lengths of individual time series, due to the short lifetimes of the magnetic features, hampered the estimate of the mean duration of these packets, as well as the resolution of the low frequency part of the spectra. \\
We have also demonstrated that at least the $v_{h}$ power spectra associated to the longest lived magnetic elements are compatible with those of random walks. \citet{2011A&A...531L...9M} have shown that once small magnetic elements reach the intergranular lanes they are subject to the buffeting action at the hands of the surrounding granular and integranular turbulent flows and they  follow random walks. However, a lot of photospheric magnetic elements may have ballistic trajectories before entering the intergranular lanes.
Additionally, we used the coherence to investigate the interaction between the transverse velocity and the longitudinal perturbations. 
From the analysis we have found that in most of the cases a high coherence is found in the high frequency band ($P < 200$ s). 
This demonstrates that, at high frequency, the vertical and the horizontal velocity are strongly coupled and interacting. This interaction is also characterized by a phase lag. The statistical analysis has shown, in fact, that the phase difference $\phi(v_{z}-v_{h})$ is found to be $\pm 90^{\circ}$. It is not possible to solve the ambiguity which is inherent to our methodology. Note that only those cases in which the coherence was above a very high threshold ($0.8$) were considered, thus strengthening the reliability of the results. \\
This result demonstrates that buffeting-induced displacements of the flux tubes are accompanied by, and possibly excite, longitudinal and compressive MHD waves.  Jafarzadeh et al. (in prep.) have found, using multi-wavelength observations, obtained from Sunrise/SuFI, that small magnetic elements are mainly dominated by fast MHD waves with velocity of the order of $40$ km/s, as well as kink waves with velocities in the same range. Both kinds of waves were upward propagating. This scenario is also supported by numerical simulations \citep{2012A&A...538A..79N}. On the other hand, \citet{2012ApJ...746..183J} have also detected slow upward propagating longitudinal waves in small magnetic elements, using high cadence broad band 2D data, visible as periodic intensity fluctuations in the range $110-600$ s. \\
\citet{lrsp-2005-3} have reviewed in detail the excitation of waves in a magnetic cylinder. They have shown, by means of numerical simulations, that kink waves and longitudinal compressive waves may coexist in the same flux tube, the latter being excited by the horizontal motion of the flux tube itself (see for example their Fig. 4 and the associated movie\footnote{The movie of the simulation is available online at the following link: \url{http://solarphysics.livingreviews.org/Articles/lrsp-2005-3/resources/cylindermodes/fast_m1_b0_running/java_movieplayer_trans.html} }). In particular, the kink mode excites a longitudinal density perturbation, whose maximum is reached at the inversion points of the flux tube's horizontal motion (i.e. when the horizontal velocity is zero). The longitudinal perturbations and the horizontal velocity are therefore $90^{\circ}$ out of phase. Although our results are consistent with this scenario, we believe that the phase shift alone is not enough to uniquely identify the mode of oscillation. This is because, the vertical velocity in both the observations and the realistic MHD simulations used, is probably the result of a mixture of modes, whose superposition hinders their identification.\\
The interaction and, therefore, the energy exchange between the kink mode and the longitudinal mode takes place over a wide range of frequencies (see Fig. \ref{phase} panel b), being more probable at high frequencies ($P<100$ s). At the frequencies of interaction, the power spectra of both kink oscillations and longitudinal oscillations show less power than the low-frequency band.
It is worth stressing that, transverse perturbations and the longitudinal mode can interact even where the power is below the maximum power achieved in the power spectrum associated with each time series. In other words, in order for the two modes to interact, it is not necessary to have a large amount of power at the same time and at the same frequency in both processes.
Although the amount of energy per unit frequency is small in this case, Fig. \ref{phase} demonstrates that this interaction can take place over a wide range of frequencies (possibly even at frequencies above the Nyquist frequency imposed by the cadence of our observations), thus the total amount of energy involved could be significantly larger.\\
These results strongly rely on the coherence and the phase estimates. 
Unfortunately, while for the power it is straightforward to give an estimate of the confidence level of its peaks in the wavelet diagram, this is not true for the phase. The major problem comes from the fact that, in the case of pure noise, there is no reference value for the angle, the latter being a variable uniformly distributed from $-\pi$ to $+\pi$. For this reason, we have provided a statistical check of our phase estimate, by exploiting the large number of magnetic elements at our disposal. This not only proves the reliability of the phase relation between $v_{z}$ and $v_{h}$, but also underlines the general character of this result, which holds for a significant fraction of magnetic elements.\\
Moreover, we have also checked whether this phase relation is found only when considering the velocity field inside the magnetic features or, rather, it is obtained also when the vertical velocity outside the magnetic elements ($v_{LNM}$) is considered. This check has demonstrated that the $\pm 90^{\circ}$ phase can be only obtained in the first case. This basically means two important things. Firstly, our phase estimate is not affected by contamination through the acoustic field surrounding the magnetic elements. Secondly, the results seen so far are closely related to the MHD waves present in small magnetic features and are not significantly affected by spurious effects coming from the external velocity field. This means that we are really taking advantage of the high spatial resolution provided by SUNRISE, which allows us to isolate the oscillation field inside small-scale magnetic elements. \\
We compared the observational results with radiative-MHD simulations, and we found them in excellent agreement. This demonstrates 
the reliability of the IMaX results and puts our findings on a firmer ground.\\
The phase histograms could suffer from the presence of a bias due to the inability to track magnetic features for a long time. This is most likely due to the short lifetimes of the magnetic features, although we cannot rule out detection problems in some cases (e.g. when a magnetic feature is lost or confused due to interactions with other features). Therefore, longer lived magnetic features could host a lot more power in longer-period oscillations.  
From the theoretical point of view, the presence of detectable longitudinal velocity oscillations inside the flux tubes, associated with buffeting-induced transverse perturbations, is in agreement with the theoretical work by \cite{Hasan2003} and \cite{Musielak2003}. These authors computed that the horizontal displacement of the flux tubes, due to the external forcing by granular buffeting, not only generates transverse oscillations, but also longitudinal oscillations, which should then be observable inside the magnetic structures as Doppler velocity oscillations. This scenario is in good agreement with our findings. Moreover, our results give new insight into the interaction between longitudinal and transverse perturbations  and, as far as we know, constitutes the first observational evidence of the interaction between these two in solar magnetic features. To this regard, one way to distinguish random horizontal motions of flux tubes from (propagating) kink waves would be to look at different heights simultaneously. In this case, for not vanishing phase lags this would indicate the presence of propagating waves. One may think that if a tube is displaced by convection in the deeper layers in one horizontal direction, it will move first in the lower layers then in the upper layers, mimicking a propagating wave. However, such a motion may set up a propagating kink wave. \\

\section{Conclusions}
In this paper we have reported the main properties of MHD waves and perturbations in small-scale magnetic elements by comparing high-resolution SUNRISE/IMaX observations to MHD simulations. This analysis reveals the interaction between transverse perturbations and longitudinal velocity oscillations in small-scale magnetic elements.\\ 
We find small-scale magnetic features in the solar photosphere to host a rich variety of MHD perturbations which we studied by means of a statistical approach. Our analysis has demonstrated that most of the magnetic elements observed display transversal oscillations with amplitudes of the order of $1$ km/s, and with a power spectrum which extends to very high frequencies ($\nu > 10$ mHz). The longitudinal waves found in the data are characterized by a lower frequency ($\nu < 7-8$ mHz). Although their power spectra present peaks at different positions, the interaction between kink displacements and longitudinal waves is, with high confidence level, found to take place in the range $P<200$ s. 
In particular we found that the Doppler velocity signal within the magnetic features shows a preference for $\pm 90^{\circ}$ phase shift with respect to the horizontal velocity of the magnetic feature itself.\\
This result is only obtained if the velocity signal inside the magnetic elements is considered, thus ruling out the contamination from the external non-magnetic oscillatory field.\\
These results are compliant with the theoretical framework in which kink waves, generated by the granular buffeting, are accompanied by longitudinal waves which are generated through non-linear interactions.\\
Small scale magnetic fields cover a significant fraction of the solar surface, therefore the presence of this kind of MHD waves and their energy exchange, may assume a remarkable importance in the energy transportation toward the upper layers of the Sun's atmosphere. 
This work has profited greatly from the high resolution and the absence of seeing characterizing the SUNRISE data, which make them ideal for studying time-dependent phenomena. New high spatial resolution and simultaneous multi-line spectropolarimetric observations of the solar quiet photosphere and chromosphere are needed to reveal clues about the propagation of these MHD waves in small scale flux tubes. \\
Furthermore, our results emphasize the need for very high temporal cadence ($< 30$ s) data in order to resolve the very high frequency part of the power spectrum.

\begin{acknowledgements}
We thank Lluis Bellot Rubio for providing the SIR inversions of the IMaX data. We also thank Laurent Gizon, Hannah Schunker and Aaron Birch for useful discussions. M. S. thanks Francesco Berrilli for useful discussions and comments. The German contribution to Sunrise is funded by the Bundesministerium für Wirtschaft und Technologie through Deutsches Zentrum für Luft- und Raumfahrt e.V. (DLR), Grant No. 50 OU 0401, and by the Innovationsfond of the President of the Max Planck Society (MPG). The Spanish contribution has been funded by the Spanish MICINN under projects ESP2006-13030-C06 and AYA2009-14105-C06 (including European FEDER funds). The HAO contribution was partly funded through NASA grant NNX08AH38G. This work has been partly supported by the WCU grant (No R31-10016) funded by the Korean Ministry of Education, Science and Technology.\\
\end{acknowledgements}

\bibliographystyle{aa}
\bibliography{lib}

\begin{thebibliography}{45}
\expandafter\ifx\csname natexlab\endcsname\relax\def\natexlab#1{#1}\fi

\bibitem[{Barthol {et~al.}(2010)Barthol, Gandorfer, Solanki, Sch\"{u}ssler,
  Chares, Curdt, Deutsch, Feller, Germerott, Grauf, Heerlein, Hirzberger,
  Kolleck, Meller, M\"{u}ller, Riethm\"{u}ller, Tomasch, Kn\"{o}lker, Lites,
  Card, Elmore, Fox, Lecinski, Nelson, Summers, Watt, {Mart\'{\i}nez Pillet},
  Bonet, Schmidt, Berkefeld, Title, Domingo, {Gasent Blesa}, {Toro Iniesta},
  {L\'{o}pez Jim\'{e}nez}, \'{A}lvarez Herrero, Sabau-Graziati, Widani,
  Haberler, H\"{a}rtel, Kampf, Levin, {P\'{e}rez Grande}, Sanz-Andr\'{e}s, \&
  Schmidt}]{Barthol2010}
Barthol, P., Gandorfer, A., Solanki, S.~K., {et~al.} 2010, Sol. Phys., 268, 1

\bibitem[{Blackman \& Tukey(1958)}]{blackman1958measurement}
Blackman, R.~B. \& Tukey, J.~W. 1958, {The measurement of power spectra from
  the point of view of communications} (Dover Publications)

\bibitem[{{Bloomfield} {et~al.}(2004){Bloomfield}, {McAteer}, {Lites}, {Judge},
  {Mathioudakis}, \& {Keenan}}]{2004ApJ...617..623B}
{Bloomfield}, D.~S., {McAteer}, R.~T.~J., {Lites}, B.~W., {et~al.} 2004, ApJ,
  617, 623

\bibitem[{{Bonet} {et~al.}(2012){Bonet}, {Cabello}, \& {S{\'a}nchez
  Almeida}}]{2012A&A...539A...6B}
{Bonet}, J.~A., {Cabello}, I., \& {S{\'a}nchez Almeida}, J. 2012, \aap, 539, A6

\bibitem[{{Cameron} {et~al.}(2007){Cameron}, {Sch{\"u}ssler}, {V{\"o}gler}, \&
  {Zakharov}}]{2007A&A...474..261C}
{Cameron}, R., {Sch{\"u}ssler}, M., {V{\"o}gler}, A., \& {Zakharov}, V. 2007,
  \aap, 474, 261

\bibitem[{Centeno {et~al.}(2009)Centeno, Collados, \& {Trujillo
  Bueno}}]{2009ApJ...692.1211C}
Centeno, R., Collados, M., \& {Trujillo Bueno}, J. 2009, $\backslash$apj, 692,
  1211

\bibitem[{{Cheung} {et~al.}(2007){Cheung}, {Sch{\"u}ssler}, \&
  {Moreno-Insertis}}]{2007A&A...467..703C}
{Cheung}, M.~C.~M., {Sch{\"u}ssler}, M., \& {Moreno-Insertis}, F. 2007, \aap,
  467, 703

\bibitem[{{De Pontieu} {et~al.}(2004){De Pontieu}, Erd\'{e}lyi, \&
  James}]{DePontieu2004}
{De Pontieu}, B., Erd\'{e}lyi, R., \& James, S.~P. 2004, Nature, 430, 536

\bibitem[{{DeForest} {et~al.}(2007){DeForest}, {Hagenaar}, {Lamb}, {Parnell},
  \& {Welsch}}]{2007ApJ...666..576D}
{DeForest}, C.~E., {Hagenaar}, H.~J., {Lamb}, D.~A., {Parnell}, C.~E., \&
  {Welsch}, B.~T. 2007, \apj, 666, 576

\bibitem[{Edwin \& Roberts(1983)}]{Edwin1983}
Edwin, P. \& Roberts, B. 1983, Sol. Phys., 88, 179

\bibitem[{Fedun {et~al.}(2011)Fedun, Shelyag, \& Erd\'{e}lyi}]{Fedun2011}
Fedun, V., Shelyag, S., \& Erd\'{e}lyi, R. 2011, ApJ, 727, 17

\bibitem[{Guglielmino {et~al.}(2012)Guglielmino, Pillet, Bonet, del
  Toro~Iniesta, Rubio, Solanki, Schmidt, Gandorfer, Barthol, \&
  Knölker}]{0004-637X-745-2-160}
Guglielmino, S.~L., Pillet, V.~M., Bonet, J.~A., {et~al.} 2012, ApJ, 745, 160

\bibitem[{Hasan {et~al.}(2003)Hasan, Kalkofen, van Ballegooijen, \&
  Ulmschneider}]{Hasan2003}
Hasan, S.~S., Kalkofen, W., van Ballegooijen, A.~A., \& Ulmschneider, P. 2003,
  ApJ, 585, 1138

\bibitem[{{Jafarzadeh} {et~al.}(2013){Jafarzadeh}, {Solanki}, {Feller}, {Lagg},
  {Pietarila}, {Danilovic}, {Riethm{\"u}ller}, \& {Mart{\'{\i}}nez
  Pillet}}]{2013A&A...549A.116J}
{Jafarzadeh}, S., {Solanki}, S.~K., {Feller}, A., {et~al.} 2013, \aap, 549,
  A116

\bibitem[{{Jess} {et~al.}(2009){Jess}, {Mathioudakis}, {Erd{\'e}lyi},
  {Crockett}, {Keenan}, \& {Christian}}]{2009Jess}
{Jess}, D.~B., {Mathioudakis}, M., {Erd{\'e}lyi}, R., {et~al.} 2009, Science,
  323, 1582

\bibitem[{{Jess} {et~al.}(2012{\natexlab{a}}){Jess}, {Pascoe}, {Christian},
  {Mathioudakis}, {Keys}, \& {Keenan}}]{2012ApJ...744L...5J}
{Jess}, D.~B., {Pascoe}, D.~J., {Christian}, D.~J., {et~al.}
  2012{\natexlab{a}}, \apjl, 744, L5

\bibitem[{{Jess} {et~al.}(2012{\natexlab{b}}){Jess}, {Shelyag}, {Mathioudakis},
  {Keys}, {Christian}, \& {Keenan}}]{2012ApJ...746..183J}
{Jess}, D.~B., {Shelyag}, S., {Mathioudakis}, M., {et~al.} 2012{\natexlab{b}},
  ApJ, 746, 183

\bibitem[{{Keller} {et~al.}(2004){Keller}, {Sch{\"u}ssler}, {V{\"o}gler}, \&
  {Zakharov}}]{2004ApJ...607L..59K}
{Keller}, C.~U., {Sch{\"u}ssler}, M., {V{\"o}gler}, A., \& {Zakharov}, V. 2004,
  \apjl, 607, L59

\bibitem[{Khomenko {et~al.}(2008{\natexlab{a}})Khomenko, Centeno, Collados, \&
  {Trujillo Bueno}}]{2008ApJ...676L..85K}
Khomenko, E., Centeno, R., Collados, M., \& {Trujillo Bueno}, J.
  2008{\natexlab{a}}, \apjl, 676, L85

\bibitem[{Khomenko {et~al.}(2008{\natexlab{b}})Khomenko, Collados, \&
  Felipe}]{Khomenko2008}
Khomenko, E., Collados, M., \& Felipe, T. 2008{\natexlab{b}}, Sol. Phys., 251,
  589

\bibitem[{{Lagg} {et~al.}(2010){Lagg}, {Solanki}, {Riethm{\"u}ller},
  {Mart{\'{\i}}nez Pillet}, {Sch{\"u}ssler}, {Hirzberger}, {Feller}, {Borrero},
  {Schmidt}, {del Toro Iniesta}, {Bonet}, {Barthol}, {Berkefeld}, {Domingo},
  {Gandorfer}, {Kn{\"o}lker}, \& {Title}}]{2010ApJ...723L.164L}
{Lagg}, A., {Solanki}, S.~K., {Riethm{\"u}ller}, T.~L., {et~al.} 2010, \apjl,
  723, L164

\bibitem[{{Manso Sainz} {et~al.}(2011){Manso Sainz}, {Mart{\'{\i}}nez
  Gonz{\'a}lez}, \& {Asensio Ramos}}]{2011A&A...531L...9M}
{Manso Sainz}, R., {Mart{\'{\i}}nez Gonz{\'a}lez}, M.~J., \& {Asensio Ramos},
  A. 2011, \aap, 531, L9

\bibitem[{{Mart{\'{\i}}nez Gonz{\'a}lez} {et~al.}(2011){Mart{\'{\i}}nez
  Gonz{\'a}lez}, {Asensio Ramos}, {Manso Sainz}, {Khomenko}, {Mart{\'{\i}}nez
  Pillet}, {Solanki}, {L{\'o}pez Ariste}, {Schmidt}, {Barthol}, \&
  {Gandorfer}}]{2011ApJ...730L..37M}
{Mart{\'{\i}}nez Gonz{\'a}lez}, M.~J., {Asensio Ramos}, A., {Manso Sainz}, R.,
  {et~al.} 2011, \apjl, 730, L37

\bibitem[{{Mart{\'{\i}}nez Gonz{\'a}lez} {et~al.}(2012){Mart{\'{\i}}nez
  Gonz{\'a}lez}, {Bellot Rubio}, {Solanki}, {Mart{\'{\i}}nez Pillet}, {Del Toro
  Iniesta}, {Barthol}, \& {Schmidt}}]{2012ApJ...758L..40M}
{Mart{\'{\i}}nez Gonz{\'a}lez}, M.~J., {Bellot Rubio}, L.~R., {Solanki}, S.~K.,
  {et~al.} 2012, \apjl, 758, L40

\bibitem[{{Mart{\'{\i}}nez Pillet} {et~al.}(2011){Mart{\'{\i}}nez Pillet}, {Del
  Toro Iniesta}, {{\'A}lvarez-Herrero}, {Domingo}, {Bonet}, {Gonz{\'a}lez
  Fern{\'a}ndez}, {L{\'o}pez Jim{\'e}nez}, {Pastor}, {Gasent Blesa}, {Mellado},
  {Piqueras}, {Aparicio}, {Balaguer}, {Ballesteros}, {Belenguer}, {Bellot
  Rubio}, {Berkefeld}, {Collados}, {Deutsch}, {Feller}, {Girela}, {Grauf},
  {Heredero}, {Herranz}, {Jer{\'o}nimo}, {Laguna}, {Meller}, {Men{\'e}ndez},
  {Morales}, {Orozco Su{\'a}rez}, {Ramos}, {Reina}, {Ramos},
  {Rodr{\'{\i}}guez}, {S{\'a}nchez}, {Uribe-Patarroyo}, {Barthol}, {Gandorfer},
  {Knoelker}, {Schmidt}, {Solanki}, \& {Vargas
  Dom{\'{\i}}nguez}}]{2011SoPh..268...57M}
{Mart{\'{\i}}nez Pillet}, V., {Del Toro Iniesta}, J.~C., {{\'A}lvarez-Herrero},
  A., {et~al.} 2011, \solphys, 268, 57

\bibitem[{Musielak {et~al.}(1989)Musielak, Rosner, \&
  Ulmschneider}]{Musielak1989}
Musielak, Z.~E., Rosner, R., \& Ulmschneider, P. 1989, ApJ, 337, 470

\bibitem[{Musielak \& Ulmschneider(2003{\natexlab{a}})}]{Musielak2003a}
Musielak, Z.~E. \& Ulmschneider, P. 2003{\natexlab{a}}, A\&A, 400, 1057

\bibitem[{Musielak \& Ulmschneider(2003{\natexlab{b}})}]{Musielak2003}
Musielak, Z.~E. \& Ulmschneider, P. 2003{\natexlab{b}}, A\&A, 406, 725

\bibitem[{Nakariakov \& Verwichte(2005)}]{lrsp-2005-3}
Nakariakov, V.~M. \& Verwichte, E. 2005, Liv. Rev. Solar Phys., 2

\bibitem[{{Narain} \& {Ulmschneider}(1996)}]{1996SSRv...75..453N}
{Narain}, U. \& {Ulmschneider}, P. 1996, \ssr, 75, 453

\bibitem[{{Nutto} {et~al.}(2012){Nutto}, {Steiner}, {Schaffenberger}, \&
  {Roth}}]{2012A&A...538A..79N}
{Nutto}, C., {Steiner}, O., {Schaffenberger}, W., \& {Roth}, M. 2012, \aap,
  538, A79

\bibitem[{{Rempel} {et~al.}(2009){Rempel}, {Sch{\"u}ssler}, {Cameron}, \&
  {Kn{\"o}lker}}]{2009Sci...325..171R}
{Rempel}, M., {Sch{\"u}ssler}, M., {Cameron}, R.~H., \& {Kn{\"o}lker}, M. 2009,
  Science, 325, 171

\bibitem[{Roberts(1983)}]{Roberts1983}
Roberts, B. 1983, Sol. Phys., 87, 77

\bibitem[{{Roberts} \& {Webb}(1978)}]{1978SoPh...56....5R}
{Roberts}, B. \& {Webb}, A.~R. 1978, \solphys, 56, 5

\bibitem[{{Roth} {et~al.}(2010){Roth}, {Franz}, {Bello Gonz{\'a}lez},
  {Mart{\'{\i}}nez Pillet}, {Bonet}, {Gandorfer}, {Barthol}, {Solanki},
  {Berkefeld}, {Schmidt}, {del Toro Iniesta}, {Domingo}, \&
  {Kn{\"o}lker}}]{2010ApJ...723L.175R}
{Roth}, M., {Franz}, M., {Bello Gonz{\'a}lez}, N., {et~al.} 2010, \apjl, 723,
  L175

\bibitem[{{Ruiz Cobo} \& {del Toro Iniesta}(1992)}]{1992ApJ...398..375R}
{Ruiz Cobo}, B. \& {del Toro Iniesta}, J.~C. 1992, ApJ, 398, 375

\bibitem[{{Solanki}(1986)}]{1986A&A...168..311S}
{Solanki}, S.~K. 1986, \aap, 168, 311

\bibitem[{{Solanki} {et~al.}(2010){Solanki}, {Barthol}, {Danilovic}, {Feller},
  {Gandorfer}, {Hirzberger}, {Riethm{\"u}ller}, {Sch{\"u}ssler}, {Bonet},
  {Mart{\'{\i}}nez Pillet}, {del Toro Iniesta}, {Domingo}, {Palacios},
  {Kn{\"o}lker}, {Bello Gonz{\'a}lez}, {Berkefeld}, {Franz}, {Schmidt}, \&
  {Title}}]{2010ApJ...723L.127S}
{Solanki}, S.~K., {Barthol}, P., {Danilovic}, S., {et~al.} 2010, \apjl, 723,
  L127

\bibitem[{{Solanki} {et~al.}(2006){Solanki}, {Inhester}, \&
  {Sch{\"u}ssler}}]{2006RPPh...69..563S}
{Solanki}, S.~K., {Inhester}, B., \& {Sch{\"u}ssler}, M. 2006, Rep. Prog. in
  Phys., 69, 563

\bibitem[{Steiner {et~al.}(1998)Steiner, Grossmann-Doerth, Knoelker, \&
  Sch{\"u}ssler}]{1998ApJ...495..468S}
Steiner, O., Grossmann-Doerth, U., Knoelker, M., \& Sch{\"u}ssler, M. 1998,
  ApJ, 495, 468

\bibitem[{Torrence \& Webster(1999)}]{Torrence1999}
Torrence, C. \& Webster, P.~J. 1999, Journal of Climate, 12, 2679

\bibitem[{{V{\"o}gler} \& {Sch{\"u}ssler}(2007)}]{2007A&A...465L..43V}
{V{\"o}gler}, A. \& {Sch{\"u}ssler}, M. 2007, \aap, 465, L43

\bibitem[{{V{\"o}gler} {et~al.}(2005){V{\"o}gler}, {Shelyag}, {Sch{\"u}ssler},
  {Cattaneo}, {Emonet}, \& {Linde}}]{2005A&A...429..335V}
{V{\"o}gler}, A., {Shelyag}, S., {Sch{\"u}ssler}, M., {et~al.} 2005, \aap, 429,
  335

\bibitem[{{Volkmer} {et~al.}(1995){Volkmer}, {Kneer}, \&
  {Bendlin}}]{1995A&A...304L...1V}
{Volkmer}, R., {Kneer}, F., \& {Bendlin}, C. 1995, \aap, 304, L1

\bibitem[{Welsch \& Longcope(2003)}]{Welsch2003}
Welsch, B.~T. \& Longcope, D.~W. 2003, ApJ, 588, 620

\end{thebibliography}

\end{document}